\documentclass[a4paper,11pt]{article}
\pdfoutput=1

\usepackage {amsmath}
\usepackage {amsfonts}
\usepackage {dsfont}
\usepackage {amsthm}
\usepackage {mathrsfs}
\usepackage {upgreek}
\usepackage {verbatim}
\usepackage{feynmp}
\usepackage{slashed}
\usepackage{subfig}
\usepackage{pstool}
\usepackage{mcite}
\usepackage{layout}

\newcommand{\BigO}[1]{\ensuremath{\operatorname{\mathcal{O}}\left(#1\right)}}

\def\FermionHeavyLoop#1{
\begin{fmfgraph}(40,40)
\fmfsurroundn{e}{#1}
\begin{fmffor}{n}{1}{1}{#1}
\fmf{dots}{e[n],i[n]}
\fmfdot{i[n]}
\end{fmffor}
\fmfcyclen{heavy,tension=#1/8}{i}{#1}
\end{fmfgraph}}

\def\FermionHeavyLoopContracted#1#2#3{
\begin{fmfgraph}(40,40)
\fmfsurroundn{e}{#1}
\begin{fmffor}{n}{1}{1}{#2-1}
\fmf{dots}{e[n],i[n]}
\end{fmffor}
\fmf{phantom}{e[#2],i[#2]}
\begin{fmffor}{n}{#2+1}{1}{#3-1}
\fmf{dots}{e[n],i[n]}
\end{fmffor}
\fmf{phantom}{e[#3],i[#3]}
\begin{fmffor}{n}{#3+1}{1}{#1}
\fmf{dots}{e[n],i[n]}
\end{fmffor}
\begin{fmffor}{n}{1}{1}{#1}
\fmfdot{i[n]}
\end{fmffor}
\fmf{dbl_dots,right=1.4,tension=0}{i[#2],i[#3]}
\fmfcyclen{heavy,tension=#1/8}{i}{#1}
\end{fmfgraph}}

\def\ScalarHeavyDoubleLoop#1{
\begin{fmfgraph}(40,40)
\fmfsurroundn{e}{2*#1}
\begin{fmffor}{n}{1}{1}{#1}
\fmf{dots}{e[2*n-1],i[n]}
\fmf{dots}{e[2*n],i[n]}
\fmfdot{i[n]}
\end{fmffor}
\fmfcyclen{dbl_dots,tension=#1/8}{i}{#1}
\end{fmfgraph}}

\def\ScalarHeavyDoubleLoopContracted#1#2#3{
\begin{fmfgraph}(40,40)
\fmfsurroundn{e}{2*#1}
\begin{fmffor}{n}{1}{1}{#2-1}
\fmf{dots}{e[2*n-1],i[n]}
\fmf{dots}{e[2*n],i[n]}
\end{fmffor}
\fmf{phantom}{e[2*#2],i[#2]}
\begin{fmffor}{n}{#2+1}{1}{#3-1}
\fmf{dots}{e[2*n-1],i[n]}
\fmf{dots}{e[2*n],i[n]}
\end{fmffor}
\fmf{phantom}{e[2*#3-1],i[#3]}
\fmf{dots}{e[2*#3],i[#3]}
\begin{fmffor}{n}{#3+1}{1}{#1}
\fmf{dots}{e[2*n-1],i[n]}
\fmf{dots}{e[2*n],i[n]}
\end{fmffor}
\fmf{dots}{e[2*#2-1],i[#2]}
\begin{fmffor}{n}{1}{1}{#1}
\fmfdot{i[n]}
\end{fmffor}
\fmf{dbl_dots,right=1.3,tension=0}{i[#2],i[#3]}
\fmf{phantom,right,tension=0}{i[#3],i[#2]}
\fmfcyclen{dbl_dots,tension=#1/8}{i}{#1}
\end{fmfgraph}}

\usepackage[hyperfootnotes=false,pdftex,bookmarks,colorlinks=false,linkbordercolor={1 1 0},urlbordercolor={1 0 0}]{hyperref}

\numberwithin{equation}{section}

\title{Fine-tuning and vacuum stability in Wilsonian effective action}
\author{Tomasz Krajewski\footnote{Tomasz.Krajewski@fuw.edu.pl}{} \; Zygmunt Lalak\footnote{Zygmunt.Lalak@fuw.edu.pl}{} \\ 
Institute of
Theoretical Physics, Faculty of Physics, University of Warsaw\\ ul. Pasteura 5,
Warsaw, Poland} 
\date{}
\begin{document}
\maketitle
\begin{abstract}
We have computed Wilsonian effective action in a~simple model with spontaneously broken chiral parity. 
We have computed Wilsonian running of relevant parameters which makes it possible to discuss in a~consistent manner the issues of fine-tuning and stability of the scalar potential. This has been compared with the standard picture based on \mbox{Gell-Mann--Low} running. Since Wilsonian running includes automatically integration of heavy degrees of freedom, the running differs markedly from the \mbox{Gell-Mann--Low} version. However, similar behaviour can be observed: scalar mass squared parameter and the quartic coupling can change sign from a~positive to a~negative one due to running which causes spontaneous symmetry breaking or an instability in the renormalizable part of the potential for a~given range of scales. However, care must be taken when drawing conclusions, because of the truncation of higher dimension operators. Taking scalar field's amplitude near the cut-off $\Lambda$ may cancel suppression due to the scale and only suppression due to small couplings partially justifies truncation in this region. Also, when taking the cut-off higher, to include larger amplitudes of the fields, higher-order irrelevant operators, whose coefficients grow with scale, may affect the conclusion about stability. The \mbox{Gell-Mann--Low} running allows one to resume relatively easily a~class of operators corresponding to large logarithms to form RGE improved effective potential valid over a~huge range of scales. In the Wilsonian approach this would correspond to following the running of a~large number of irrelevant operators, which is technically problematic.
As for the issue of fine-tuning, since in the Wilsonian approach power-law terms are not subtracted, one can clearly observe the quadratic sensitivity of fine-tuning measure to the change of the cut-off scale. The Wilsonian version of the radiative symmetry breaking mechanism has been described.  
\end{abstract}
\section{Introduction}
The recent discovery of the Higgs boson at the Large Hadron Collider \cite{Aad:2012tfa,Chatrchyan:2012ufa} promotes the question about protection of the electroweak breaking scale to one of the most puzzling problems of fundamental physics. The observed compatibility of properties of the newly observed particle with predictions coming from Standard Model additionally strengthens tension between standard theoretical reasoning which results in prediction of new physics near the electroweak scale and reality. Neither supersymmetry nor composite Higgs models, perhaps most attractive solutions to hierarchy problem, are favoured by the observed value of Higgs mass \cite{Giardino:2012ww,Carmi:2012in,Plehn:2012iz}. Moreover, production and decay rates  have not provided unambiguous evidence for new physics.

This situation strengthens the need of revisiting the naturalness principle which have been used as a~guide for model building, since its formulation in the late 1970s and early 1980s \cite{Susskind:1978ms,'tHooft:1979bh,Veltman:1980mj}. Numerous authors \cite{Aoki:2012xs,Farina:2013mla,Jegerlehner:2013cta,Jegerlehner:2013nna,Bian:2013xra,deGouvea:2014xba,Bar-Shalom:2014taa} propose new definitions of naturalness. Our goal is less ambitious. We shall try to state clearly a~treatment of fine-tuning
 based on Wilsonian effective action and corresponding Wilsonian renormalization group. Idea of Wilsonian effective action is close to the intuitive understanding of cutoff regularization. In standard discussion based on quadratic divergences the artificial meaning of a~scale of effective theory is given to the regularization parameter $\Lambda$. This effects in the regularization dependence of this kind of analysis. On the contrary, in the Wilsonian method high energy modes are integrated out in a~self consistent, regularization independent way and effective theory has a~well-defined effective action. Moreover, this treatment is universal and depends very weakly on a~preferred UV completion (quantum gravity, string theory, etc.). Main impact on effective action from states with masses greater than the scale of the effective theory can be parametrized by values of couplings of the Wilsonian effective action. Further corrections are highly suppressed as far as heavy masses are separated from the scale of the effective theory. 

Given a~model where vacuum expectation value of a~scalar field can be generated with quantum corrections we can also show how the stability of the effective action looks like from the point of view of Wilsonian running. This has been compared with the standard picture based on \mbox{Gell-Mann--Low} running. Since Wilsonian running includes automatically integration of heavy degrees of freedom, the running differs markedly from the \mbox{Gell-Mann--Low} version. Nevertheless, similar behaviour can be observed: scalar mass squared parameter and the quartic coupling can change sign from a~positive to a~negative one due to running. This causes spontaneous symmetry breaking or an instability in the renormalizable part of the potential for a~given range of scales. However, care must be taken when drawing conclusions, because of the truncation of higher dimension operators.

While simple cut-off analysis of scalar field models has been performed earlier, the goal of the present note is to use consistently Wilsonian approach and to make clear comparison with the discussion based on the Gell-Mann-Low running.

The paper is organized as follows. In Section \ref{model} we specify the model and define truncation. We argue in Subsection \ref{Lagrangian} that this model should present behaviour similar to that known from the SM. In Subsection \ref{truncation} we show in what sense RGEs for chosen truncation can be thought as an~analogue of 1-loop \mbox{Gell-Mann--Low} running. In Section \ref{RGE} we present calculated RGEs. Their numerical solution is discussed in Section \ref{numerical}. Section \ref{fine_tuning} is dedicated to numerical estimation of fine-tuning of parameters of Wilsonian effective action. In Section \ref{stability} we discuss the issue of radiative stability of the effective action and in \ref{conclusion} we summarize our results. Appendix \ref{Wilsonian} contains brief introduction to Wilsonian RGE and Functional Renormalization Group (FRG) methods. The derivatives of loop integrals used during calculations are given in Appendix \ref{loop_integrals}. In Appendix \ref{matching} we discuss matching conditions which give parameters of Wilsonian effective action in terms of measurable quantities. 

\section{Basic features of the model\label{model}}
\subsection{Couplings \label{Lagrangian}}
For the sake of clarity we consider a~simple model that exhibits certain interesting features of the SM. The model consists of a~massless Dirac fermion $\Psi$ which couples via Yukawa interaction to a~real scalar field $\Phi$ with a~quartic self-coupling. 
This Lagrangian takes the form:
\begin{equation}
\mathcal{L}=i \overline{\Psi} \slashed{\partial} \Psi + \frac{1}{2} \partial_\mu \Phi \partial^\mu \Phi - \frac{1}{2} M^2 \Phi^2 - Y \Phi \overline{\Psi} \Psi - \frac{\lambda}{4!} {\Phi}^4. \label{Lagrangian_density}
\end{equation}
The above Lagrangian is symmetric under (chiral) $\mathbb{Z}_2$ which acts on $\Phi$ as $\Phi\to-\Phi$ and on $\Psi$ as $\Psi \to \gamma^5 \Psi$. We consider the case of non-zero vacuum expectation value for the field $\Phi$, which breaks this symmetry spontaneously. In broken symmetry phase the Lagrangian density \eqref{Lagrangian_density} will take the form:
\begin{equation}
\mathcal{L}=i \overline{\Psi} \slashed{\partial} \Psi - m \overline{\Psi} \Psi + \frac{1}{2} \partial_\mu \Phi \partial^\mu \Phi - \frac{1}{2} M^2 \Phi^2 - Y \Phi \overline{\Psi} \Psi -\frac{g}{3!} \Phi^3 - \frac{\lambda}{4!} {\Phi}^4. \label{lagriangian_density_broken}
\end{equation}
The fermion $\Psi$ allows one to model top quark coupling to Higgs boson, which is known to give the main contribution to quadratic divergences in the mass of the SM scalar and to high-scale instability of the quartic coupling. 
%The massless $\Psi$ can obtain a~mass from the vev of $\Phi$. We use Dirac fermion instead of a~Weyl fermion in order to avoid introducing chiral formalism for the latter. However, all results can be easily extended to such a~situation. Gauge interactions have been neglected to avoid additional complications in calculating Wilsonian RGE posed by gauge invariance. 

This model was previously investigated with methods of FRG in \cite{Clark:1992jr,Clark:1994ya,Gies:2013fua} in order to estimate non-perturbative bound on Higgs boson mass. The same issue was discussed in \cite{Gies:2014xha} with a~slightly different Lagrangian. In \cite{Branchina:2005tu} stability of potential was discussed with the help of the naive cut-off procedure. 

\subsection{Perturbative derivation of RGE\label{truncation}}
We have calculated Wilsonian renormalization group equations at the lowest non-trivial order. The Wilsonian action can include an infinite number of non-renormalizable operators, however they are suppressed at the low cut-off scale. Hence our truncation contains the following operators:
\begin{equation}
\mathcal{L}_{\Lambda}=i \overline{\Psi}_{\Lambda} \slashed{\partial} \Psi_{\Lambda} + \frac{1}{2} \partial_\mu \Phi_{\Lambda} \partial^\mu \Phi_{\Lambda} - \frac{1}{2} M^2_{\Lambda} {\Phi_{\Lambda}}^2 - Y_{\Lambda} \Phi_{\Lambda} \overline{\Psi}_{\Lambda} \Psi_{\Lambda} - \frac{\lambda_{\Lambda}}{4!} {\Phi_{\Lambda}}^4. \label{truncation_Lagrangian}
\end{equation}
We define Wilson coefficients $M^2_{\Lambda}$, $Y_{\Lambda}$ and $\lambda_{\Lambda}$ as values of respectively two-, three- and four- point Green's functions at the kinematic point with vanishing external momenta. We use following graphics for cutoff propagator of the scalar field $\Phi$
\begin{equation}
\parbox{20mm}{
 \begin{fmffile}{heavy_scalar_propagator}
   \fmfset{arrow_len}{3mm}
  \fmfset{thick}{thin}
 \begin{fmfgraph*}(20,10)
 \fmfleft{i}
 \fmfright{o}
 \fmf{dbl_dots}{i,o}
 \end{fmfgraph*}
 \end{fmffile}
 } = \frac{\theta_0(p^2-\Lambda^2)}{p^2+M_\Lambda^2}
 \end{equation}
 and the fermionic field $\Psi$
 \begin{equation}
\parbox{20mm}{
 \begin{fmffile}{heavy_fermion_propagator}
   \fmfset{arrow_len}{3mm}
  \fmfset{thick}{thin}
 \begin{fmfgraph*}(20,10)
 \fmfleft{i}
 \fmfright{o}
 \fmf{heavy}{i,o}
 \end{fmfgraph*}
 \end{fmffile}
 } = \frac{\theta_0(p^2-\Lambda^2)}{\slashed{p}+m_\Lambda}.
\end{equation}

In the diagrams lines
 \begin{fmffile}{light_scalar_propagator}
 \begin{fmfgraph}(20,10)
   \fmfset{arrow_len}{3mm}
  \fmfset{thick}{thin}
 \fmfleft{i}
 \fmfright{o}
 \fmf{dots}{i,o}
 \end{fmfgraph}
 \end{fmffile}
and 
 \begin{fmffile}{light_fermion_propagator}
 \begin{fmfgraph}(20,10)
   \fmfset{arrow_len}{3mm}
  \fmfset{thick}{thin}
 \fmfleft{i}
 \fmfright{o}
 \fmf{fermion}{i,o}
 \end{fmfgraph}
 \end{fmffile}
represent respectively scalar and fermionic low-energy modes.
\subsection{Lowest order calculation}
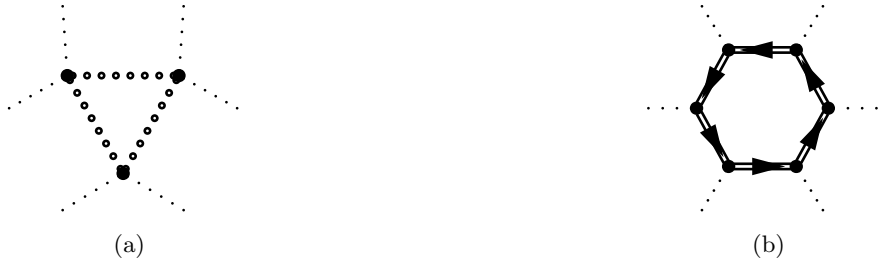
\begin{figure}[!h]
\begin{minipage}{.5\linewidth}
\centering
\subfloat[]{\label{scalar_loop_dim_6}
\begin{fmffile}{scalar_loop}
  \fmfset{arrow_len}{3mm}
  \fmfset{thick}{thin}
\ScalarHeavyDoubleLoop{3}
\end{fmffile}
}
\end{minipage}
\begin{minipage}{.5\linewidth}
\centering
\subfloat[]{\label{fermion_loop_dim_6}
\begin{fmffile}{fermion_loop} 
  \fmfset{arrow_len}{3mm}
  \fmfset{thick}{thin}
 \FermionHeavyLoop{6}
 \end{fmffile}
}
\end{minipage}
\caption{Lowest order diagrams which contribute to generating $\Phi^6$ operator in effective Lagrangian density coming from bare Lagrangian density \eqref{Lagrangian_density}.}
\label{loop_dim_6}
\end{figure}

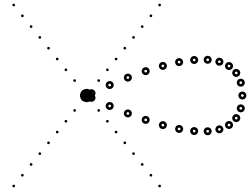
\begin{figure}[!h]
\centering
 \begin{fmffile}{dim_6}
   \fmfset{arrow_len}{3mm}
  \fmfset{thick}{thin}
 \begin{fmfgraph}(32,32)
 \fmfleft{i1,i2}
 \fmfright{o1,o2}
 \fmf{dots}{i1,v}
 \fmf{dots}{i2,v}
 \fmf{dots}{v,o1}
 \fmf{dots}{v,o2}
 \fmf{dbl_dots,tension=1/2}{v,v}
 \fmfdot{v}
 \end{fmfgraph}
 \end{fmffile}
 \caption{Lowest order contribution to $\beta$-function for coupling $\lambda$ from $\Phi^6$ operator.\label{beta_dim_6}}
 \end{figure}

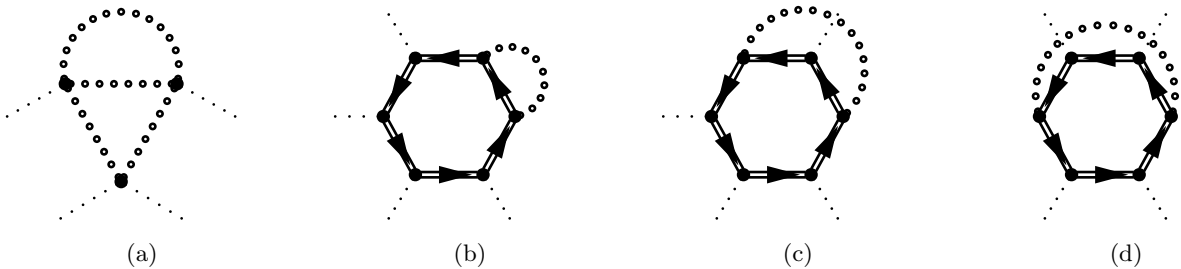
\begin{figure}[h!]
\centering
\subfloat[]{\label{scalar_loop_contracted_dim_6_12}
 \begin{fmffile}{scalar_loop_contracted_12} 
   \fmfset{arrow_len}{3mm}
  \fmfset{thick}{thin}
 \ScalarHeavyDoubleLoopContracted{3}{1}{2}
 \end{fmffile}
 }
\subfloat[]{\label{fermion_loop_contracted_dim_6_12}
 \begin{fmffile}{fermion_loop_contracted_12} 
  \fmfset{arrow_len}{3mm}
  \fmfset{thick}{thin}
 \FermionHeavyLoopContracted{6}{1}{2}
 \end{fmffile}
 }
\subfloat[]{\label{fermion_loop_contracted_dim_6_13}
 \begin{fmffile}{fermion_loop_contracted_13} 
   \fmfset{arrow_len}{3mm}
  \fmfset{thick}{thin}
 \FermionHeavyLoopContracted{6}{1}{3}
 \end{fmffile}
 }
\subfloat[]{\label{fermion_loop_contracted_dim_6_14}
 \begin{fmffile}{fermion_loop_contracted_14} 
   \fmfset{arrow_len}{3mm}
  \fmfset{thick}{thin}
 \FermionHeavyLoopContracted{6}{1}{4}
 \end{fmffile}
 }
\caption{2-loop contribution to \mbox{Gell-Mann--Low} type running of $\lambda$ which appear in Wilsonian running as a~contribution from $\Phi^6$ and higher dimension operators.\label{full_dim_6}}
\end{figure}

The RGE that we obtain can be thought of as an analogue of 1-loop \mbox{Gell-Mann--Low} type RGE. To see the relation more precisely, let us for a~moment assume that we add an operator $\Phi^6$ to truncation \eqref{truncation_Lagrangian}. Its lowest-order contribution to $\beta$-function for $\lambda$ comes from diagrams with loop propagator starting and ending in the same vertex. For a~theory with bare Lagrangian of the form \eqref{Lagrangian_density} the operator $\Phi^6$ must be, in standard treatment, generated by loop diagrams. Diagrams given in Fig. \ref{scalar_loop_dim_6} and Fig. \ref{fermion_loop_dim_6} show the lowest order contribution to Wilsonian coefficient of the $\Phi^6$ operator.
These contributions are respectively proportional to ${\lambda}^3$ and ${Y}^6$. Lowest order contribution to $\beta$-function for the coupling $\lambda$ given by the $\Phi^6$ operator comes from loop diagram shown in Fig. \ref{beta_dim_6}.
 Combining diagrams from Figs. \ref{loop_dim_6} and \ref{beta_dim_6} we obtain 2-loop diagrams given in Fig. \ref{full_dim_6}.
To sum up, operator $\Phi^6$ is generated at one loop level and the second loop is needed to obtain contribution to the $\beta$-function for $\lambda$. The lowest order non-trivial contributions to \mbox{Gell-Mann--Low} type $\beta$-function are of the order ${\lambda}^2$ and ${Y}^4$. The contribution to Wilsonian RGE generated by the operator $\Phi^6$ appears at the 2-loop level in the \mbox{Gell-Mann--Low} type RGE.

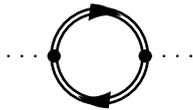
\begin{figure}[!h]
\centering
 \begin{fmffile}{Z1-fermion-fermion}
   \fmfset{arrow_len}{3mm}
  \fmfset{thick}{thin}
 \begin{fmfgraph}(32,20)
 \fmfleft{i}
 \fmfright{o}
 \fmf{dots}{i,v1}
 \fmf{fermion,left,tension=2/3}{v1,v2}
 \fmf{dots}{v2,o} 
 \fmf{fermion,left,tension=0}{v2,v1}
 \fmfdot{v1,v2}
 \end{fmfgraph}
 \end{fmffile}
 \caption{Fermion loop diagram that gives momentum-dependent contribution to scalar self-energy.\label{momentum_dependent}}
 \end{figure}

On the other hand, we neglected all higher dimension operators with derivatives, for example $\partial_\mu \Phi \partial^\mu \Phi \Phi^2$.
It is an easy task to check that the diagram from the Fig. \ref{momentum_dependent} gives momentum-dependent\footnote{Diagram from the Fig. \ref{momentum_dependent} depends on external momentum of a~scalar if we integrate over all modes or just a~part of fermionic modes in the loop.} contribution to scalar self-energy.
More precisely, contribution coming from this diagram depends on momentum squared $p^2$ logarithmically. To recover this dependence from the effective action one needs to consider an infinite number of operators of the form $\Box^n \Phi^2 \colon n=2, \dots$. All these operators are irrelevant near the Gaussian fixed point (the free theory).
\section{Flow equations\label{RGE}}

\subsection{Calculating RGE using Mathematica}

Our procedure of calculating $\beta$ functions is as follows:
\begin{enumerate}
\item Draw all 1PI and counterterms diagrams with certain number of external fields and write down formal expressions for them.
\item Expand loop integrals in series of external momenta and choose interesting terms (for vertex corrections we set external momenta to zero).
\item Represent loop integrals as standard Passarino--Veltman \cite{Passarino:1978jh} functions.
\item Express Passarino--Veltman functions in terms of functions $I_N$ with IR cutoff $\Lambda$ introduced in \cite{Bilal:2007ne}.
\item Differentiate result with respect to $\Lambda$. This step gives expressions in terms of derivatives of $I_N$ functions.
\item Substitute derivatives of $I_N$ functions by expressions given in Appendix \ref{loop_integrals}.
\end{enumerate}
We have assumed such a~procedure for two reasons. Firstly, by calculating effective action before differentiation with respect to the cutoff $\Lambda$ we avoid problems connected with a~sharp cutoff derivative (however, one must perform detailed calculation of derivatives with respect to external momenta). Secondly, this algorithm is easy to implement using FeynArts \cite{Hahn:2000kx} and FeynCalc \cite{Mertig:1990an}.
FeynArts can easily generate 1-loop 1PI diagrams for the Lagrangian density, which we use as an input for FeynRules \cite{Alloul:2013bka}. FeynCalc without any modification of the source code calculates Passarino--Veltman representation of diagrams generated by FeynArts. Finally we substitute (using Mathematica package modelled after ANT package \cite{Angel:2013hla}) Passarino--Veltman integrals by their derivatives with respect to cutoff $\Lambda$ which we calculated in advance. Calculations performed in Mathematica provide a~crosscheck for calculations made by hand. 

Using this producer differences between Wilsonian effective action treatment and standard one based on cut-off regularization are easy to observe and origin of terms in Wilsonian RGE coming from different diagrams can be determined. 

\subsection{RGE for the model \label{RGE_eq}}
As we discuss in the Appendix \ref{flow_equations} it is convenient to express the Wilsonian RGE in terms of dimensionless parameters. We use dimensionless parameters $\nu_{\Lambda}:=\frac{v_{\Lambda}}{{\Lambda}}$, ${\Omega_{\Lambda}}^2:=\frac{{M_{\Lambda}}^2}{{\Lambda}^2}$, $\omega_{\Lambda}:=\frac{m_{\Lambda}}{{\Lambda}}$ and $\gamma_\Lambda:= \frac{g_\Lambda}{{\Lambda}}$. We define $v_{\Lambda}$ by the requirement that the shift $\Phi\mapsto\Phi-v_{\Lambda}$ gives effective action without $\mathbb{Z}_2$-odd terms.

Derivatives with respect to the scale $\Lambda$ of the diagrams from Figs. \ref{v_broken}, \ref{Z1_broken}, \ref{Z2_broken}, \ref{y_broken}, \ref{g3_broken} and \ref{lambda_broken} give flow equations respectively for vacuum expectation value of $\Phi$ field, masses of scalar $\Phi$ and fermion $\Psi$, Yukawa coupling, Wilson coefficient $g$ for the operator $\Phi^3$ and the coupling $\lambda$. In addition we determine the scaling of fields from diagrams given in Figs. \ref{Z1_broken} and \ref{Z2_broken}. The RGE read as follows:
\begin{figure}[h!]
\centering
\subfloat[]{\label{tadpole-fermion}
 \begin{fmffile}{tadpole-fermion}
   \fmfset{arrow_len}{3mm}
  \fmfset{thick}{thin}
 \begin{fmfgraph}(32,20)
 \fmfleft{i}
 \fmfright{o}
 \fmf{heavy,right,tension=2/3,left}{i,v}
 \fmf{heavy,right,tension=2/3,left}{v,i}
 \fmf{dots,tension=1}{v,o}
 \fmfdot{v}
 \end{fmfgraph}
 \end{fmffile}
 }
\subfloat[]{\label{tadpole-scalar}
 \begin{fmffile}{tadpole-scalar}
   \fmfset{arrow_len}{3mm}
  \fmfset{thick}{thin}
 \begin{fmfgraph}(32,20)
 \fmfleft{i}
 \fmfright{o}
 \fmf{dbl_dots,right,tension=2/3,left}{i,v}
 \fmf{dbl_dots,right,tension=2/3,left}{v,i}
 \fmf{dots,tension=1}{v,o}
 \fmfdot{v}
 \end{fmfgraph}
 \end{fmffile}
 }
\caption{Tadpole diagrams which give contribution to running of $v_{\Lambda}$ parameter. \label{v_broken}}
\end{figure}
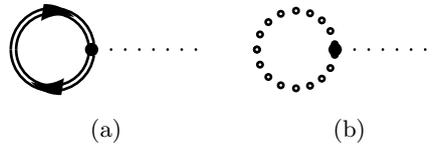
\begin{figure}[!h]
\centering
 \begin{fmffile}{Z2-fermion-scalar}
   \fmfset{arrow_len}{3mm}
  \fmfset{thick}{thin}
 \begin{fmfgraph}(32,20)
 \fmfleft{i}
 \fmfright{o}
 \fmf{fermion}{i,v1}
 \fmf{heavy,right,tension=1/2}{v1,v2}
 \fmf{fermion}{v2,o} 
 \fmf{dbl_dots,left,tension=0}{v1,v2}
 \fmfdot{v1,v2}
 \end{fmfgraph}
 \end{fmffile}
\caption{Feynman diagram which give contribution to running of the fermion $\Psi$ mass parameter $m_{\Lambda}$ and field renormalization. \label{Z2_broken}}
\end{figure}
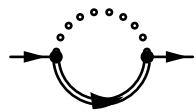
\begin{figure}[h!]
\centering
\subfloat[]{\label{Z1-fermion-fermion}
 \begin{fmffile}{Z1-fermion-fermion}
   \fmfset{arrow_len}{3mm}
  \fmfset{thick}{thin}
 \begin{fmfgraph}(32,20)
 \fmfleft{i}
 \fmfright{o}
 \fmf{dots}{i,v1}
 \fmf{heavy,left,tension=1/2}{v1,v2}
 \fmf{dots}{v2,o} 
 \fmf{heavy,left,tension=0}{v2,v1}
 \fmfdot{v1,v2}
 \end{fmfgraph}
 \end{fmffile}
 }
\subfloat[]{\label{Z1-scalar}
 \begin{fmffile}{Z1-scalar}
   \fmfset{arrow_len}{3mm}
  \fmfset{thick}{thin}
 \begin{fmfgraph}(32,20)
 \fmfleft{i}
 \fmfright{o}
 \fmf{dots}{i,v}
 \fmf{dbl_dots,right,tension=3/5}{v,v}
 \fmf{dots}{v,o} 
 \fmfdot{v}
 \end{fmfgraph}
 \end{fmffile}
 }
\subfloat[]{\label{Z1-scalar-scalar}
 \begin{fmffile}{Z1-scalar-scalar}
   \fmfset{arrow_len}{3mm}
  \fmfset{thick}{thin}
 \begin{fmfgraph}(32,20)
 \fmfleft{i}
 \fmfright{o}
 \fmf{dots}{i,v1}
 \fmf{dbl_dots,left,tension=1/2}{v1,v2}
 \fmf{dots}{v2,o} 
 \fmf{dbl_dots,left,tension=0}{v2,v1}
 \fmfdot{v1,v2}
 \end{fmfgraph}
 \end{fmffile}
 } 
 \caption{Feynman diagrams which give contribution to running of the scalar $\Phi$ mass squared parameter ${M_{\Lambda}}^2$ and field renormalization. \label{Z1_broken}} 
\end{figure}
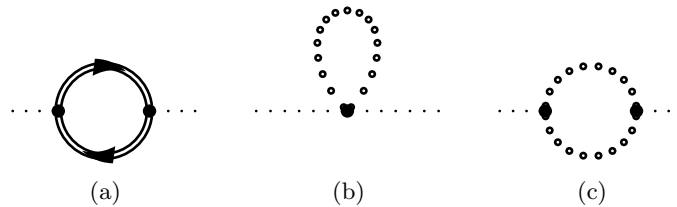
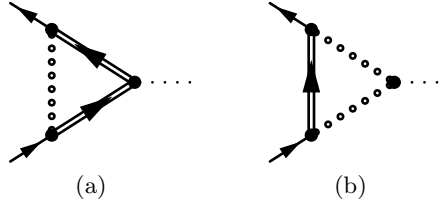
\begin{figure}[h!]
\centering
 \subfloat[]{\label{Yukawa-fermion-fermion-scalar}
 \begin{fmffile}{Yukawa-fermion-fermion-scalar}
   \fmfset{arrow_len}{3mm}
  \fmfset{thick}{thin}
 \begin{fmfgraph}(35,28)
 \fmfleft{i1,i2}
 \fmfright{o}
 \fmf{fermion,tension=2}{i1,v1}
 \fmf{fermion,tension=2}{v2,i2}
 \fmf{dbl_dots,tension=0}{v1,v2}
 \fmf{heavy,tension=1}{v1,v3}
 \fmf{heavy,tension=1}{v3,v2}
 \fmf{dots,tension=3}{v3,o}
 \fmfdot{v1,v2,v3}
 \end{fmfgraph}
 \end{fmffile}
 } 
 \subfloat[]{\label{Yukawa-scalar-scalar-fermion}
 \begin{fmffile}{Yukawa-scalar-scalar-fermion}
   \fmfset{arrow_len}{3mm}
  \fmfset{thick}{thin}
 \begin{fmfgraph}(35,28)
 \fmfleft{i1,i2}
 \fmfright{o}
 \fmf{fermion,tension=2}{i1,v1}
 \fmf{fermion,tension=2}{v2,i2}
 \fmf{heavy,tension=0}{v1,v2}
 \fmf{dbl_dots,tension=1}{v1,v3}
 \fmf{dbl_dots,tension=1}{v3,v2}
 \fmf{dots,tension=3}{v3,o}
 \fmfdot{v1,v2,v3}
 \end{fmfgraph}
 \end{fmffile}
 }
 \caption{Feynman diagrams which give contribution to running of the Yukawa coupling $Y_{\Lambda}$. \label{y_broken}}
\end{figure}
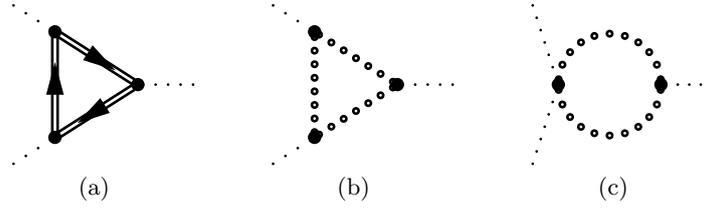
\begin{figure}[h!]
\centering
 \subfloat[]{\label{g3-fermion-fermion-fermion}
 \begin{fmffile}{g3-fermion-fermion-fermion}
   \fmfset{arrow_len}{3mm}
  \fmfset{thick}{thin}
 \begin{fmfgraph}(35,28)
 \fmfleft{i1,i2}
 \fmfright{o}
 \fmf{dots,tension=2}{i1,v1}
 \fmf{dots,tension=2}{v2,i2}
 \fmf{heavy,tension=0}{v1,v2}
 \fmf{heavy,tension=1}{v2,v3}
 \fmf{heavy,tension=1}{v3,v1}
 \fmf{dots,tension=3}{v3,o}
 \fmfdot{v1,v2,v3}
 \end{fmfgraph}
 \end{fmffile}
 }
 \subfloat[]{\label{g3-scalar-scalar-scalar}
 \begin{fmffile}{g3-scalar-scalar-scalar}
   \fmfset{arrow_len}{3mm}
  \fmfset{thick}{thin}
 \begin{fmfgraph}(35,28)
 \fmfleft{i1,i2}
 \fmfright{o}
 \fmf{dots,tension=2}{i1,v1}
 \fmf{dots,tension=2}{v2,i2}
 \fmf{dbl_dots,tension=0}{v1,v2}
 \fmf{dbl_dots,tension=1}{v1,v3}
 \fmf{dbl_dots,tension=1}{v3,v2}
 \fmf{dots,tension=3}{v3,o}
 \fmfdot{v1,v2,v3}
 \end{fmfgraph}
 \end{fmffile}
 }
 \subfloat[]{\label{g3-scalar-scalar}
 \begin{fmffile}{g3-scalar-scalar}
   \fmfset{arrow_len}{3mm}
  \fmfset{thick}{thin}
 \begin{fmfgraph}(35,28)
 \fmfleft{i1,i2}
 \fmfright{o}
 \fmf{dots,tension=1}{i1,v1}
 \fmf{dots,tension=1}{v1,i2}
 \fmf{dbl_dots,tension=1/4,left}{v1,v2}
 \fmf{dbl_dots,tension=1/4,left}{v2,v1}
 \fmf{dots,tension=1}{o,v2}
 \fmfdot{v1,v2}
 \end{fmfgraph}
 \end{fmffile}
 } 
\caption{Feynman diagrams which give contribution to running of the parameter ${g}_{\Lambda}$ in the ordered phase. \label{g3_broken}}
\end{figure}
\begin{figure}[h!]
\centering
 \subfloat[]{\label{lambda-scalar-scalar}
 \begin{fmffile}{lambda-scalar-scalar}
   \fmfset{arrow_len}{3mm}
  \fmfset{thick}{thin}
 \begin{fmfgraph}(35,28)
 \fmfleft{i1,i2}
 \fmfright{o1,o2}
 \fmf{dots,tension=1}{i1,v1}
 \fmf{dots,tension=1}{v1,i2}
 \fmf{dbl_dots,tension=1/4,left}{v1,v2}
 \fmf{dbl_dots,tension=1/4,left}{v2,v1}
 \fmf{dots,tension=1}{o1,v2}
 \fmf{dots,tension=1}{o2,v2}
 \fmfdot{v1,v2}
 \end{fmfgraph}
 \end{fmffile}
 }
 \subfloat[]{\label{lambda-scalar-scalar-scalar-scalar}
 \begin{fmffile}{lambda-scalar-scalar-scalar-scalar}
   \fmfset{arrow_len}{3mm}
  \fmfset{thick}{thin}
 \begin{fmfgraph}(35,28)
 \fmfleft{i1,i2}
 \fmfright{o1,o2}
 \fmf{dots,tension=1}{i1,v1}
 \fmf{dots,tension=1}{v4,i2}
 \fmf{dbl_dots,tension=0.2}{v1,v2}
 \fmf{dbl_dots,tension=0.2}{v2,v3}
 \fmf{dbl_dots,tension=0.2}{v3,v4}
 \fmf{dbl_dots,tension=0.2}{v4,v1}
 \fmf{dots,tension=1}{o1,v2}
 \fmf{dots,tension=1}{o2,v3}
 \fmfdot{v1,v2,v3,v4}
 \end{fmfgraph}
 \end{fmffile}
 }
 \subfloat[]{\label{lambda-scalar-scalar-scalar}
 \begin{fmffile}{lambda-scalar-scalar-scalar}
   \fmfset{arrow_len}{3mm}
  \fmfset{thick}{thin}
 \begin{fmfgraph}(35,28)
 \fmfleft{i1,i2}
 \fmfright{o1,o2}
 \fmf{dots,tension=1}{i1,v1}
 \fmf{dots,tension=1}{v3,i2}
 \fmf{dbl_dots,tension=1/4}{v1,v2}
 \fmf{dbl_dots,tension=1/4}{v2,v3}
 \fmf{dbl_dots,tension=1/4}{v3,v1}
 \fmf{dots,tension=1}{o1,v2}
 \fmf{dots,tension=1}{o2,v3}
 \fmfdot{v1,v2,v3}
 \end{fmfgraph}
 \end{fmffile}
 }
 \subfloat[]{\label{lambda-fermion-fermion-fermion-fermion}
 \begin{fmffile}{lambda-fermion-fermion-fermion-fermion}
   \fmfset{arrow_len}{3mm}
  \fmfset{thick}{thin}
 \begin{fmfgraph}(35,28)
 \fmfleft{i1,i2}
 \fmfright{o1,o2}
 \fmf{dots,tension=1}{i1,v1}
 \fmf{dots,tension=1}{v4,i2}
 \fmf{heavy,tension=0.2}{v1,v2}
 \fmf{heavy,tension=0.2}{v2,v3}
 \fmf{heavy,tension=0.2}{v3,v4}
 \fmf{heavy,tension=0.2}{v4,v1}
 \fmf{dots,tension=1}{o1,v2}
 \fmf{dots,tension=1}{o2,v3}
 \fmfdot{v1,v2,v3,v4}
 \end{fmfgraph}
 \end{fmffile}
 }
\caption{Feynman diagrams which give contribution to running of the Wilson coefficient $\lambda_{\Lambda}$ for the operator $\Phi^4$.\label{lambda_broken}}
\end{figure}
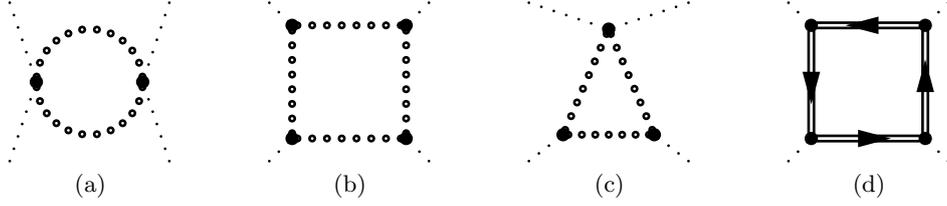
\begin{multline}
{\Lambda} \frac{d \nu_{\Lambda}}{d{\Lambda}}=-{\nu_{\Lambda}}+\frac{1}{{\Omega_{\Lambda}}^2}\left[\frac{8Y_{\Lambda}}{(4 \pi )^2} \frac{\omega_{\Lambda}}{(1+{\omega_{\Lambda}}^2)}-\frac{{{\gamma}_{\Lambda}}}{(4 \pi )^2} \frac{1}{(1+{\Omega_{\Lambda}}^2)}\right]\\
-\frac{\nu_{\Lambda}}{2}\left[ 4\frac{Y_{\Lambda}^2}{(4 \pi) ^2} \frac{\left(3-{\omega_{\Lambda}}^2\right) \left(1+4{\omega_{\Lambda}}^2\right) }{3 \left({\omega_{\Lambda}}^2+1\right)^3}{\Omega_{\Lambda}}^2+\frac{{{\gamma}_{\Lambda}}^2}{3(4 \pi) ^2}\frac{2{\Omega_{\Lambda}}^2-1}{(1+{\Omega_{\Lambda}}^2)^3}\right].
\end{multline}
\begin{multline}
{\Lambda} \frac{d \omega_{\Lambda}}{d{\Lambda}}=-{\omega_{\Lambda}}-\frac{2Y_{\Lambda}^2\omega_{\Lambda}}{(4 \pi )^2} \left[\frac{1}{(1+{\Omega_{\Lambda}}^2) \left(1+{\omega_{\Lambda}}^2\right)}-\frac{1}{(1+{\Omega_{\Lambda}}^2)
 \left({\Omega_{\Lambda}}^2-{\omega_{\Lambda}}^2\right)}\right.\\
\left.+\frac{1}{\left({\Omega_{\Lambda}}^2-{\omega_{\Lambda}}^2\right)^2}\log
 \left(\frac{1+{\Omega_{\Lambda}}^2}{1+{\omega_{\Lambda}}^2}\right)\right]
 -Y_{\Lambda} \left[\frac{8Y_{\Lambda}}{(4 \pi )^2} \frac{\omega_{\Lambda}}{(1+{\omega_{\Lambda}}^2)}-\frac{{{\gamma}_{\Lambda}}}{(4 \pi )^2} \frac{1}{(1+{\Omega_{\Lambda}}^2)})\right]
\end{multline}
\begin{multline}
{\Lambda} \frac{d {\Omega_{\Lambda}}^2}{d{\Lambda}}=-2 {\Omega_{\Lambda}}^2+4\frac{Y_{\Lambda}^2}{(4 \pi) ^2} \left[\frac{\left(3-{\omega_{\Lambda}}^2\right) \left(1+4{\omega_{\Lambda}}^2\right) }{3 \left({\omega_{\Lambda}}^2+1\right)^3}{\Omega_{\Lambda}}^2-\frac{2{\omega_{\Lambda}}^2-2} {\left({\omega_{\Lambda}}^2+1\right)^2}\right]\\
+\frac{\lambda_{\Lambda}}{(4 \pi) ^2}\frac{1}{ ({\Omega_{\Lambda}}^2+1)}+\frac{{{\gamma}_{\Lambda}}^2}{(4 \pi) ^2}\left[\frac{2}{3}\frac{1}{1+{\Omega_{\Lambda}}^2}-\frac{2}{3}\frac{1}{(1+{\Omega_{\Lambda}}^2)^2}+\frac{{\Omega_{\Lambda}}^2}{(1+{\Omega_{\Lambda}}^2)^3}\right]\\
-{\gamma}_{\Lambda} \left[\frac{8Y_{\Lambda}}{(4 \pi )^2} \frac{\omega_{\Lambda}}{(1+{\omega_{\Lambda}}^2)}-\frac{{{\gamma}_{\Lambda}}}{(4 \pi )^2} \frac{1}{(1+{\Omega_{\Lambda}}^2)})\right]
\end{multline}
\begin{multline}
{\Lambda} \frac{d Y_{\Lambda}}{d{\Lambda}}=\frac{Y_{\Lambda}^3}{(4 \pi) ^2} \left[\frac{{\omega_{\Lambda}}^2-1}{ \left({\omega_{\Lambda}}^2+1\right)^2 ({\Omega_{\Lambda}}^2+1)}-\frac{8}{3\left( \omega_{\Lambda}^2+1\right)}+\frac{38}{3 \left(\omega_{\Lambda}^2+1\right)^2}-\frac{6}{3 \left(\omega_{\Lambda}^2+1\right)^3}\right.\\
+ \left.\frac{2}{\left({\Omega_{\Lambda}}^2-{\omega_{\Lambda}}^2\right)^2} \log \left(\frac{{\Omega_{\Lambda}}^2+1}{{\omega_{\Lambda}}^2+1}\right)-\frac{2}{({\Omega_{\Lambda}}^2+1)\left({\Omega_{\Lambda}}^2-{\omega_{\Lambda}}^2\right)}\right]\\
+ \frac{{{\gamma}_{\Lambda}^2} {Y_{\Lambda}} }{(4 \pi) ^2}\left[\frac{1}{3} \frac{1}{\left({\Omega_{\Lambda}}^2+1\right)^2} -\frac{1}{2}\frac{1}{\left({\Omega_{\Lambda}}^2+1\right)^3}\right]-\frac{2{{\gamma}_{\Lambda}}Y_{\Lambda}^2}{(4 \pi) ^2}\frac{\omega_{\Lambda}}{(1+{\Omega_{\Lambda}}^2)(1+{\omega_{\Lambda}}^2)}
\end{multline}
\begin{multline}
{\Lambda} \frac{d {{\gamma}_{\Lambda}}}{d{\Lambda}}=-{\gamma}_{\Lambda}+\frac{3{{\gamma}_{\Lambda}} \lambda_{\Lambda}}{(4 \pi) ^2}\frac{1}{{\Omega_{\Lambda}}^2+1}-\frac{7{{\gamma}_{\Lambda}}^3}{2(4 \pi) ^2}\frac{1}{({\Omega_{\Lambda}}^2+1)^3}-\frac{Y_{\Lambda}^3 \omega_{\Lambda}}{(4 \pi) ^2}\frac{16\left(3-{\omega_{\Lambda}}^2\right)}{({\omega_{\Lambda}}^2+1)^3}\\
+\frac{{{\gamma}_{\Lambda}} Y_{\Lambda}^2}{(4 \pi) ^2} \frac{2 \left(3-{\omega_{\Lambda}}^2\right)(1+4{\omega_{\Lambda}}^2)}{\left({\omega_{\Lambda}}^2+1\right)^3}
-\lambda_{\Lambda} \left[\frac{8Y_{\Lambda}}{(4 \pi )^2} \frac{\omega_{\Lambda}}{(1+{\omega_{\Lambda}}^2)}-\frac{{{\gamma}_{\Lambda}}}{(4 \pi )^2} \frac{1}{(1+{\Omega_{\Lambda}}^2)})\right]
\end{multline}
\begin{multline}
{\Lambda} \frac{d \lambda_{\Lambda}}{d{\Lambda}}=3 \frac{\lambda_{\Lambda}^2}{(4 \pi )^2}\frac{1}{(1+{\Omega_{\Lambda}}^2)^2}- \frac{2 Y_{\Lambda}^4}{(4 \pi )^2}\left[\frac{1}{\left(1+{\omega_{\Lambda}}^2\right)^2}-8\frac{{\omega_{\Lambda}}^2}{\left(1+{\omega_{\Lambda}}^2\right)^4}\right]
+\frac{8}{3} \frac{\lambda_{\Lambda} Y_{\Lambda}^2 }{(4 \pi )^2} \frac{ \left(3-{\omega_{\Lambda}}^2\right)(1+4{\omega_{\Lambda}}^2)}{\left(1+{\omega_{\Lambda}}^2\right)^3}\\
+\frac{6{{\gamma}_{\Lambda}}^4}{(4 \pi )^2}\frac{1}{({\Omega_{\Lambda}}^2+1)^4}+\frac{2{{\gamma}_{\Lambda}^2} \lambda_{\Lambda}}{(4 \pi )^2}\left[\frac{2}{3}\frac{1}{({\Omega_{\Lambda}}^2+1)^2}-\frac{7}{({\Omega_{\Lambda}}^2+1)^3}\right].
\end{multline}
\section{Numerical solutions of RGE\label{numerical}} 
\begin{figure}[!h]
\includegraphics[width=\textwidth]{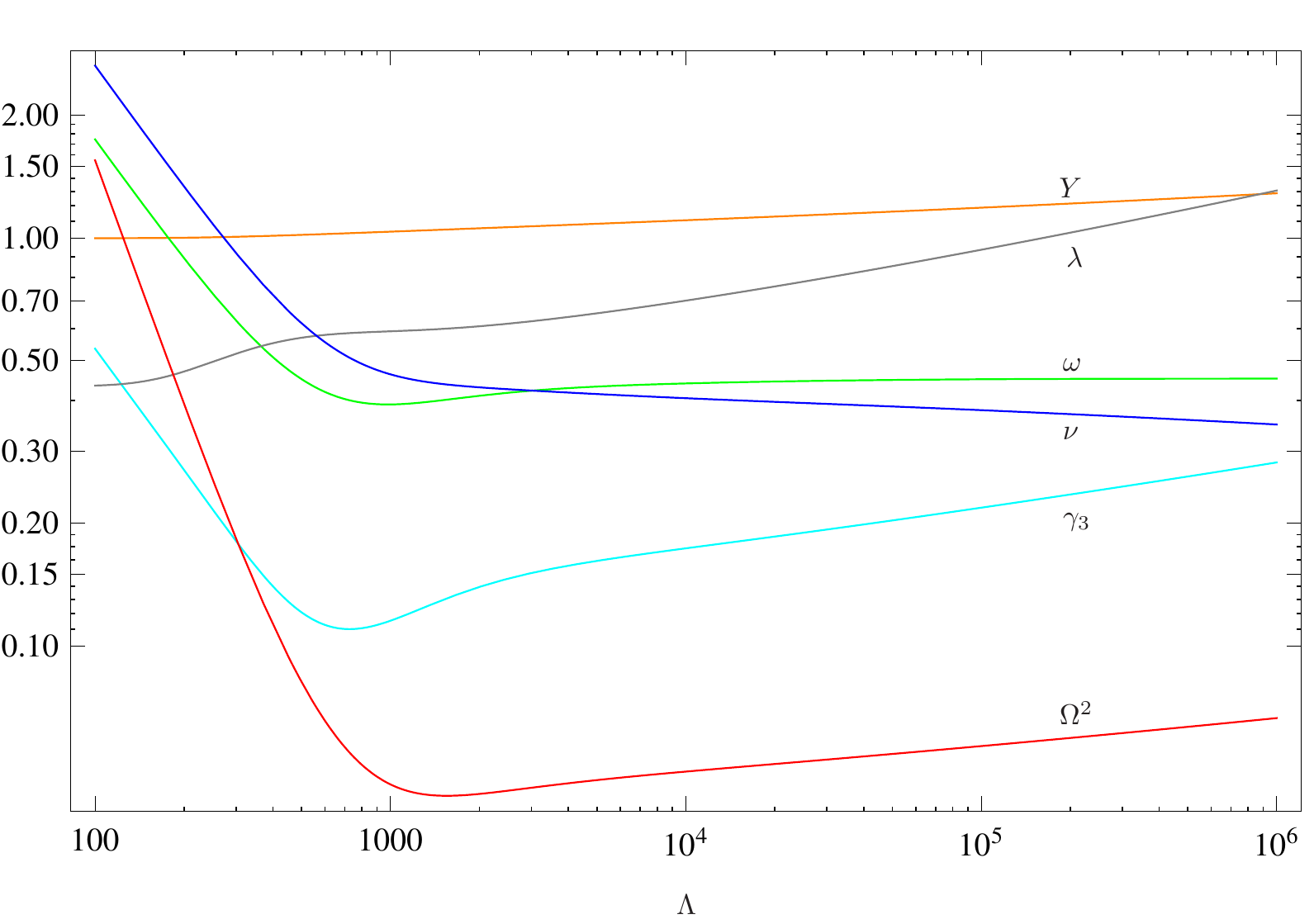}
\caption{Example of numerical solution of RGE corresponding to: $m_{ph}=174$, $M_{ph}=125$, $\lambda_{ph}=0.2$, $v_{ph}=264$, ${g}_{ph}=52.8$ and $Y_{ph}=1$.\label{numerical_solution}}
\end{figure}

The RGE \eqref{RGE_eq} have been solved numerically. We used matching conditions given in Appendix \ref{matching} to compute initial conditions for RGE at $\Lambda=100$ (we use units of GeV through the paper) in terms of physical quantities (see Appendix \ref{matching} for definitions) with renormalization scale $\mu=100$. The example solution with values: $m_{ph}=174$, $M_{ph}=125$, $\lambda_{ph}=0.2$, $v_{ph}=264$, ${g}_{ph}=52.8$ and $Y_{ph}=1$ is presented in the Fig. \ref{numerical_solution}. Double-logarithmic plot in the Fig. \ref{numerical_solution} shows parameters of the effective Wilsonian action as functions of the scale $\Lambda$. Orange line represents Yukawa coupling $Y_\Lambda$ which runs typically rather slowly. Gray line corresponds to the quartic coupling $\lambda_{\Lambda}$. This coupling runs faster, because of the contribution from the fermionic loop. Couplings $\omega_\Lambda$ (green line), $\Omega^2_\Lambda$ (red line) and ${\gamma}_\Lambda$ (cyan line) for low values of $\Lambda$ run like relevant couplings due to rescaling, but after reaching scales of the order of the masses, they change their behaviour to a~slow running near constant value. The behaviour of the above couplings for high vales of $\Lambda$ is caused by the quadratic divergencies (or more precisely by the same diagrams which generate quadratic divergences). The same behaviour is manifested by vacuum expectation value $\nu_\Lambda$ plotted as the blue line.

\begin{figure}
 \begin{minipage}{.5\linewidth}
\centering
\subfloat[]{\label{numerical_omega}
\includegraphics[width=\textwidth]{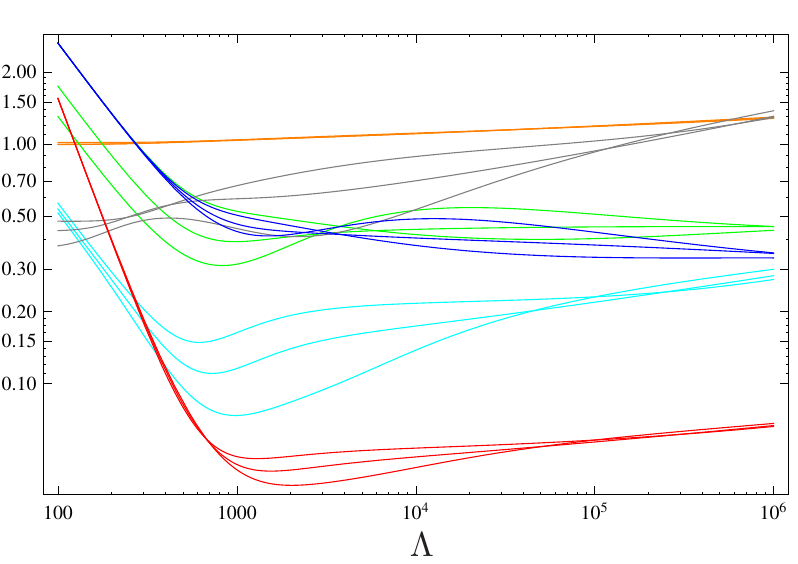}
 }
\end{minipage}
 \begin{minipage}{.5\linewidth}
\centering
\subfloat[]{\label{numerical_Omega}
\includegraphics[width=\textwidth]{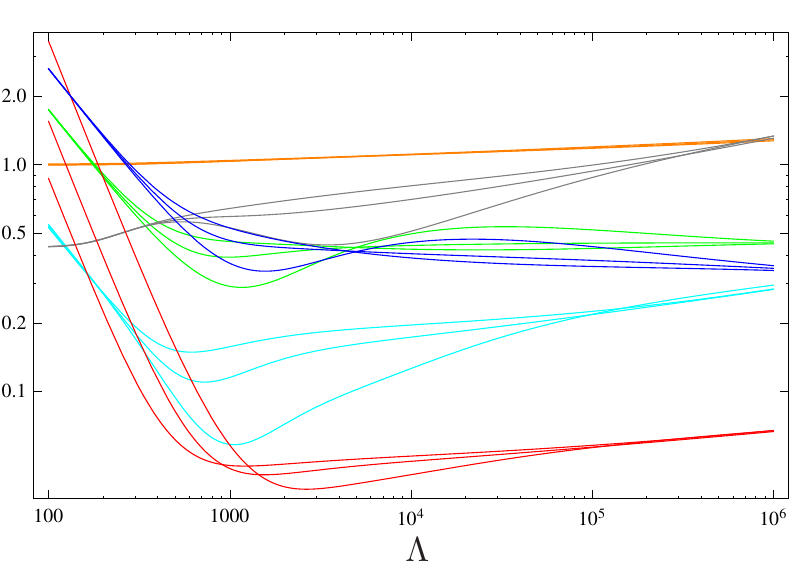}
 }
\end{minipage}
\par
 \begin{minipage}{.5\linewidth}
\centering
\subfloat[]{\label{numerical_g}
\includegraphics[width=\textwidth]{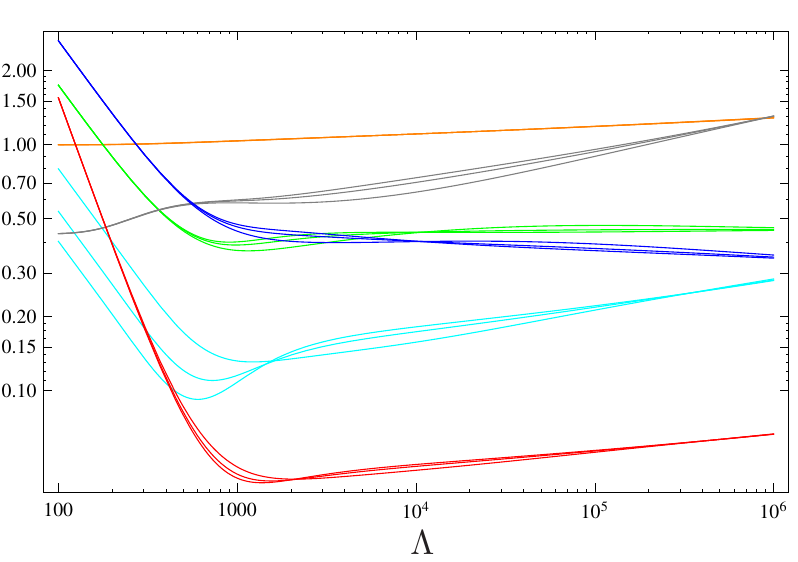}
 }
\end{minipage}
 \begin{minipage}{.5\linewidth}
\centering
\subfloat[]{\label{numerical_l}
\includegraphics[width=\textwidth]{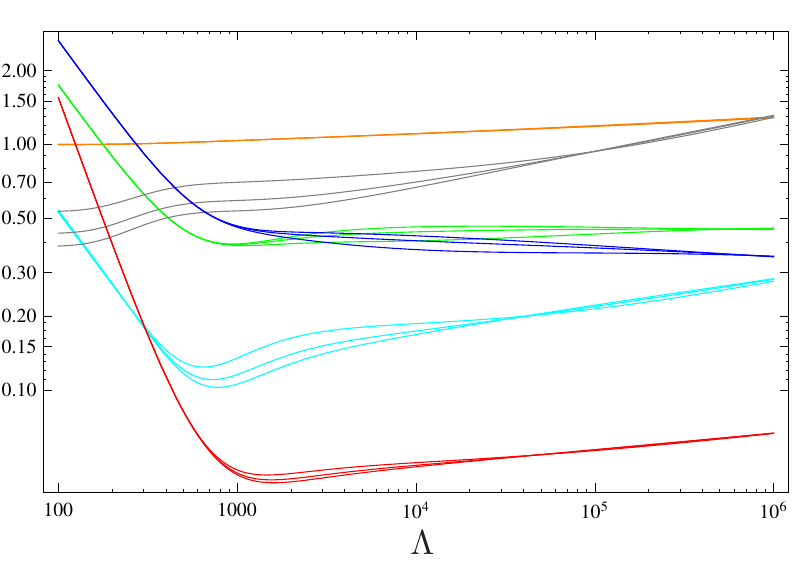}
 }
\end{minipage}
\caption{Solutions for changed initial conditions respectively: \ref{numerical_omega} $m_{ph}$, \ref{numerical_Omega} $M_{ph}$, \ref{numerical_g} ${g}_{ph}$ and \ref{numerical_l} $\lambda_{ph}$ multiplied or divided by the factor $\frac{3}{4}$.\label{numerical_variation}}
\end{figure}

Solutions with different initial conditions have the same qualitative behaviour. In the Fig. \ref{numerical_variation} we plotted families of solutions with a~single parameter varied: the solution with a~physical quantity multiplied by $\frac{3}{4}$ and another one with the same parameter multiplied by $\frac{4}{3}$. Reference solution has been plotted as well. Fig. \ref{numerical_omega} shows solutions with initial conditions $m_{ph}=\frac{3}{4}174$ and $m_{ph}=\frac{4}{3}174$. For solutions presented in Figs. \ref{numerical_Omega}, \ref{numerical_g} and \ref{numerical_l} we have respectively changed $M_{ph}$, ${g}_{ph}$ and $\lambda_{ph}$. In all plots we use the same colors as in Fig. \ref{numerical_solution} to indicate analogous parameters. 

The important observation is that all presented solutions give values very close to each other for $\Lambda$ of the order of $10^6$. Small change of parameters in effective action at high scale generates physical parameters different by orders of magnitude, since the solutions corresponding to different low-scale parameters run very closely to each other when the scale grows. This is the sign that a~fine-tuning of parameters in effective action at high scales is required in order to get the prescribed values of physical observables.

\begin{figure}[!h]
\includegraphics[width=\textwidth]{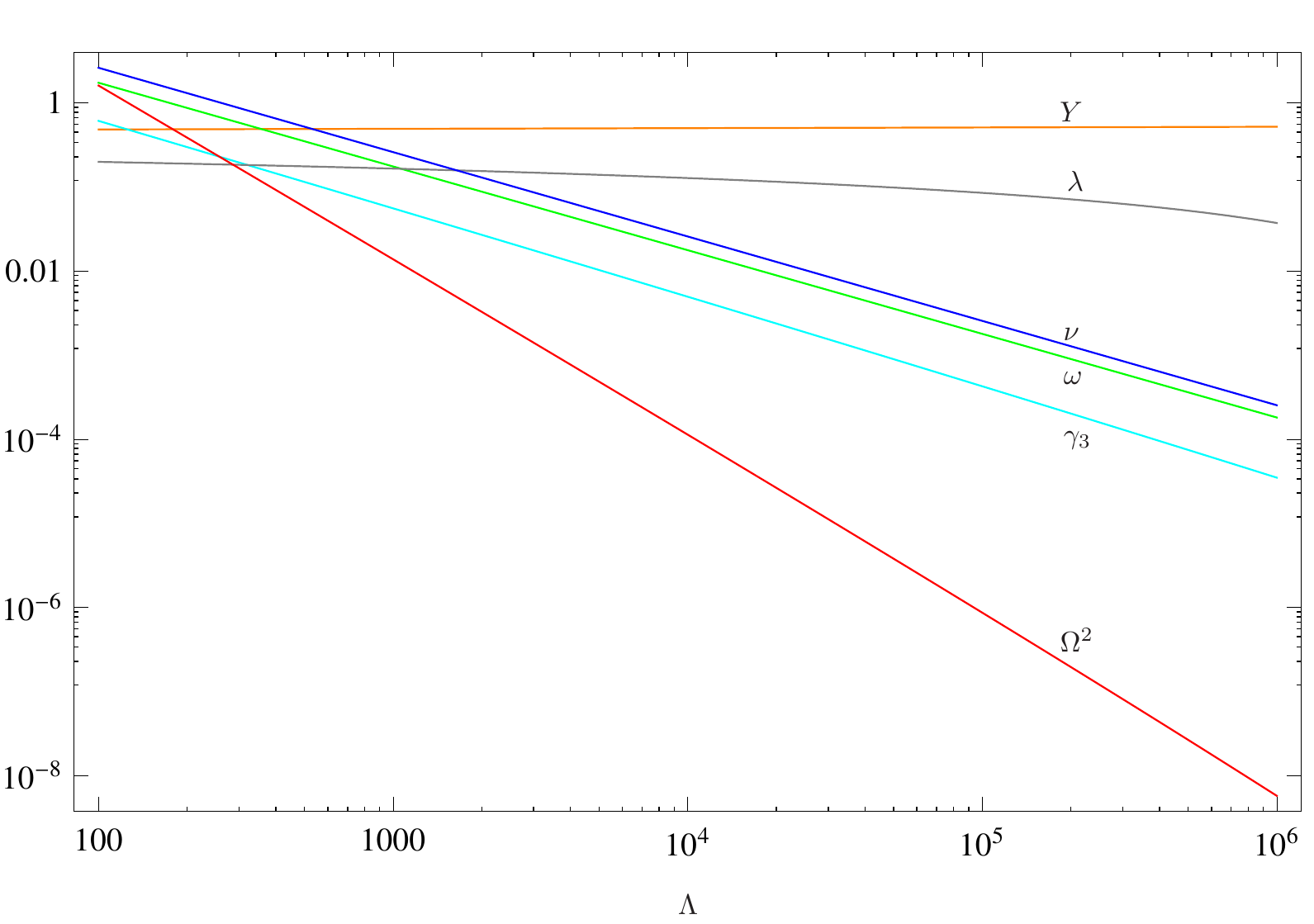}
\caption{Numerical solution of \mbox{Gell-Mann--Low} type RGE for discussed model corresponding to: $m_{ph}=174$, $M_{ph}=125$, $\lambda_{ph}=0.2$, $v_{ph}=264$, ${g}_{ph}=52.8$ and $Y_{ph}=0.5$.\label{plot_Gell-Mann--Low}}
\end{figure}
The example of numerical solution for \mbox{Gell-Mann--Low} type RGE for the same theory is given in the Fig. \ref{plot_Gell-Mann--Low}. Comparing the Fig. \ref{plot_Gell-Mann--Low} with the Fig. \ref{numerical_solution} one finds that the flow of parameters of Wilsonian effective action is much more complicated than the running in \mbox{Gell-Mann--Low} method. One should note that Wilsonian RGE accommodate decoupling of massive particles i.e. corrections from particles with masses grater than $\Lambda$ are strongly suppressed. 

\section{Fine-tuning\label{fine_tuning}} 
The standard measure $\Delta_{c_i}$ of fine-tuning with respect to the variable $c_i$ is defined as 
\begin{equation}
\Delta_{c_i} = \frac{\partial \log v^2}{\partial \log {c_i}^2},\label{finetuning_definition}
\end{equation}
where $c_i$ is a~coupling in the model and $v$ is the vacuum expectation value of the field which breaks symmetry spontaneously (here - chiral parity). 
As a~measure of fine-tuning of the whole model we take \cite{Ellis:1986yg,Barbieri:1987fn,Ghilencea:2012gz}:
\begin{equation}
\Delta = \left( \sum_i {\Delta_{c_i}}^2\right)^{\frac{1}{2}}.\label{finetuning_definition_2}
 \end{equation} 
We have computed $\Delta_{c_i}$ for parameters of the effective action as functions of scale $\Lambda$. 

\begin{figure}
\centering
\subfloat[]{\label{finetuning_omega}
\includegraphics[width=0.5 \textwidth]{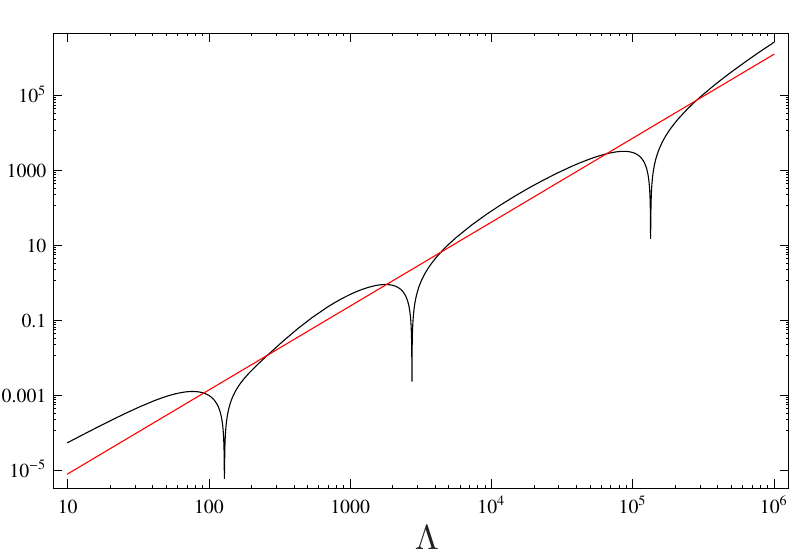}
 }
\subfloat[]{\label{finetuning_Omega}
\includegraphics[width=0.5 \textwidth]{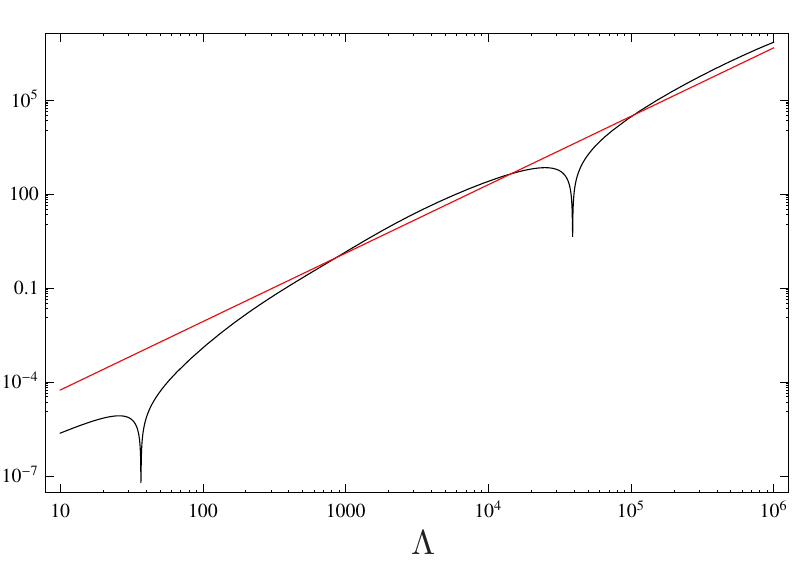}
 }
\par\vspace{-10pt}
\subfloat[]{\label{finetuning_g}
\includegraphics[width=0.5 \textwidth]{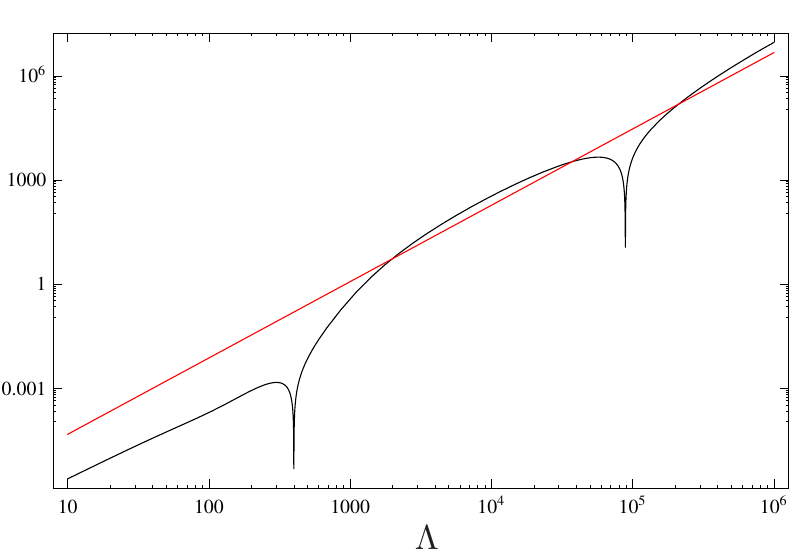}
 }
\subfloat[]{\label{finetuning_l}
\includegraphics[width=0.5 \textwidth]{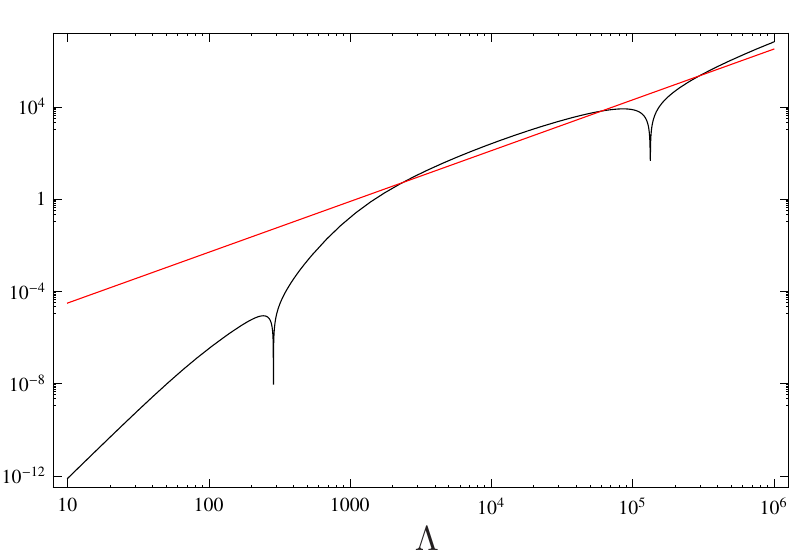}
 }
 \par\vspace{-10pt}
\subfloat[]{\label{finetuning_Y}
\includegraphics[width=0.5 \textwidth]{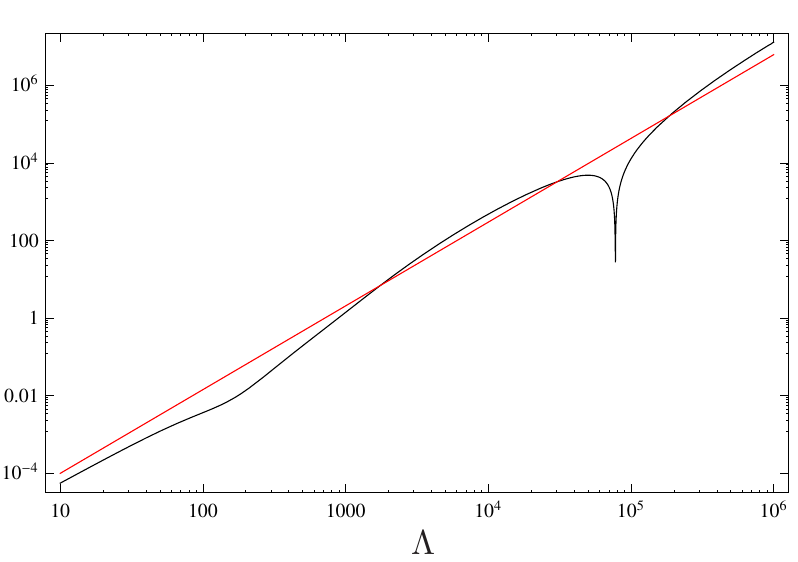}
 }
\subfloat[]{\label{finetuning_all}
\includegraphics[width=0.5 \textwidth]{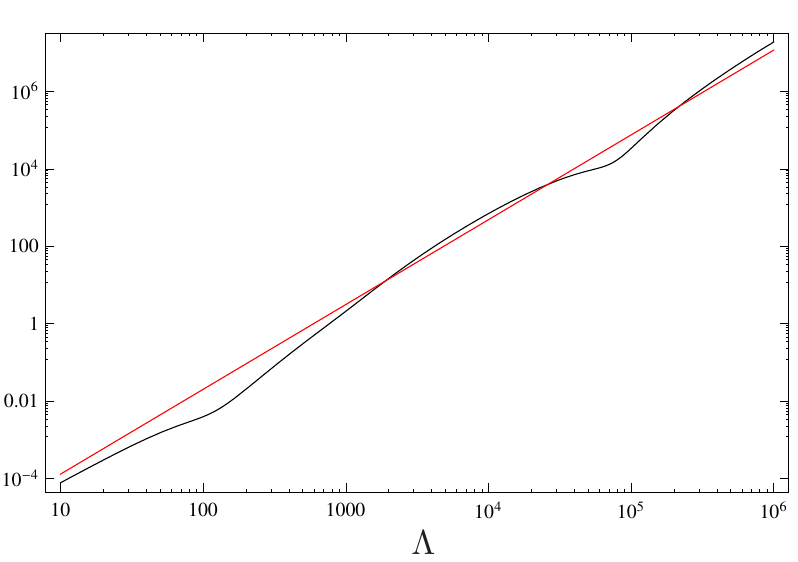}
 }
\caption{Fine-tuning measures $\Delta_{c_i}$ \eqref{finetuning_definition} for dimensionless parameters: \mbox{\protect\subref{finetuning_omega}---$\omega_\Lambda$}, \mbox{\protect\subref{finetuning_Omega}---$\Omega^2_\Lambda$}, \mbox{\protect\subref{finetuning_g}---${\gamma}_\Lambda$}, \mbox{\protect\subref{finetuning_l}---$\lambda_\Lambda$} and \mbox{\protect\subref{finetuning_Y}---$Y_\Lambda$} as functions of $\Lambda$ in Wilsonian effective action. Combined fine-tuning measure \eqref{finetuning_definition_2} is shown in plot \protect\subref{finetuning_all}. Fitted power laws are given as red lines.\protect\label{numerical_finetuning}}
\end{figure}

Unfortunately, the effective action for $\Lambda=0$ cannot be obtained by direct numerical integration of RGE. The left-hand side of RGE given in Section \ref{RGE} can be rewritten as
\begin{equation}
\Lambda \frac{d c_i}{d \Lambda} = \frac{d c_i}{d \log
\left(\Lambda/\Lambda_0\right)},
\end{equation}
where $c_i$ is a~dimensionless parameter and $\Lambda_0$ is the scale at which initial conditions are set. For the purpose of numerical integration 
couplings are functions of $t:=\log \left(\Lambda/\Lambda_0\right)$. The point $\Lambda=0$ corresponds to the limit $t \to -\infty$ which cannot be reached using numerical methods. For that reason we approximated $v_\Lambda$ for $\Lambda=0$ ($v_0$), by the value at $\Lambda=10^{-4}$ i.e. $v_{10^{-4}}$. We used $\Lambda=10^{-4}$, because this turns out to be the lowest scale which gives $\nu_\Lambda$ safe from numerical errors. On the other hand, $v_\Lambda$ changes very slowly between $\Lambda=1$ and $\Lambda=10^{-4}$, so $v_{10^{-4}}$ should be good approximation for $v_0$. To sum up, we have computed fine-tuning measure \eqref{finetuning_definition} by taking numerical derivatives of $\nu_{\Lambda}$ with respect to dimensionless parameters $\omega_\Lambda$, $\Omega^2_\Lambda$, ${\gamma}_\Lambda$, $\lambda_\Lambda$, $Y_\Lambda$ over the range of scales $10<\Lambda<10^6$. 
Figs. \ref{finetuning_omega}, \ref{finetuning_Omega}, \ref{finetuning_g}, \ref{finetuning_l} and \ref{finetuning_Y} show respectively $\left| \frac{\partial \log \nu^2_{10^{-4}}}{\partial \log{\omega_\Lambda}^2} \right|$, $\left| \frac{\partial \log \nu^2_{10^{-4}}}{\partial \log {\Omega^2_\Lambda}^2} \right|$, $\left| \frac{\partial \log \nu^2_{10^{-4}}}{\partial \log {{\gamma}_\Lambda}^2} \right|$, $\left| \frac{\partial \log \nu^2_{10^{-4}}}{\partial \log{\lambda_\Lambda}^2} \right|$ and $\left| \frac{\partial \log \nu^2_{10^{-4}}}{\partial \log {Y_\Lambda}^2} \right|$.
 
The spikes visible in the Fig. \ref{numerical_finetuning} are points where derivatives \eqref{finetuning_definition} change their signs. Due to logarithmic scale and finite resolution of plots the zeros of derivatives cannot be correctly depicted and are represented in Fig. \ref{numerical_finetuning} as a~finite spikes. Even if one of the derivatives vanishes, the others are typically non-zero and \eqref{finetuning_definition_2} stays non-zero and smooth. In Fig. \ref{finetuning_all} the measure \eqref{finetuning_definition_2} as a~function of scale $\Lambda$ is shown. The power function $\propto \Lambda^p$ which has been fitted to fine-tuning curve is shown in red in each plot in the Fig. \ref{numerical_finetuning}. The fitted powers $p$ are given in Table \ref{power}. The power-like functions have been fitted over the interval $10^{3} \leq \Lambda \leq 10^{6} $ (that is above assumed mass thresholds). The reason is the visible change of the behaviour of the flow of parameters below $10^{3}$. On the other hand, the flows above $10^{3}$ can be smoothly extrapolated to arbitrarily high scales. 
\begin{table}
\centering
\begin{tabular}{|c|c|c|c|c|c|c|}\hline
 &\multicolumn{6}{c|}{coupling}\\
 \cline{2-7}
 & $\omega_\Lambda$ & $\Omega^2_\Lambda$ & ${g}_\Lambda$ & $\lambda_\Lambda$ & $Y_\Lambda$ & combined\\
 \hline
p & 2.24 & 2.19 & 2.20 & 2.20 & 2.16 & 2.19\\ \hline
$\sigma$ & 0.05 & 0.04 & 0.04 & 0.04 & 0.04 & 0.02\\ \hline
\end{tabular}
\caption{Estimated powers $p$ and their standard deviations $\sigma$ from the fits to the fine-tuning measures \eqref{finetuning_definition}.\label{power}}
\end{table}
\section{Vacuum stability\label{stability}} 
An interesting issue is the question of spontaneous symmetry breaking and stability of the potential seen from the point of view of the Wilsonian approach. 
In this approach one starts with a~bare action at a~high scale and keeps integrating out consecutive shells of momenta, or coarse-graining, to obtain effective action at lower scales. Eventually, the vacuum structure should emerge in the infra-red limit. 

In the model studied in this paper one can try to answer the question whether the $\mathbb{Z}_2$ symmetry of \eqref{Lagrangian_density} can be broken by radiative corrections to the scalar mass parameter. What one finds is that for the low values of $\Omega_{\Lambda}^2$ and high values of Yukawa coupling $Y_{\Lambda}$ in the effective action at high $\Lambda$, scalar mass-squared parameter can flow to a negative value at low $\Lambda$. The 
change of sign of $M_{\Lambda}^2$ indicates that stable vacuum of the theory must have non-zero vacuum expectation value of scalar field $\Phi$ (as long as the quartic coupling stays positive). Moreover, quartic scalar coupling $\lambda_{\Lambda}$ can run negative for higher $\Lambda$ which shows similar behaviour as the one observed in the \mbox{Gell-Mann--Low} type running (Fig. \ref{plot_Gell-Mann--Low}) known from Standard Model. In the context of SM the zero of quartic self-coupling is usually considered as an indication of instability of the electroweak vacuum and of the existence of a~second minimum of the scalar potential. In the Wilsonian approach however, simple analysis based on quartic coupling alone is insufficient, because higher dimension operators with higher powers of the scalar field $\Phi$, which we suppressed in our truncation \eqref{truncation_Lagrangian}, may dominate scalar potential for large values of $\Phi$. The impact coming from higher dimension operators was previously investigated in \cite{Gies:2013fua,Gies:2014xha} and \cite{Lalak:2014qua}. The results presented in \cite{Gies:2013fua} validate our observation that quartic coupling constant $\lambda$ can be driven negative in UV by RGE flow. On the other hand analysis of running of higher dimension operators presented in \cite{Gies:2013fua} supports the statement that higher dimension operators stabilise scalar potential. This is at odds with the explicit calculation of \cite{Lalak:2014qua} where examples with instabilities induced by higher order operators have been given. To draw strong conclusions one needs a~procedure of resummation of possibly large contributions to scalar potential coming from operators with all higher powers of $\Phi$. 
However, the observed instability of Wilsonian quartic coupling may be seen as an indication of a~crossover behaviour at higher scales, since in the region of $\langle \Phi \rangle$ comparable to $\Lambda$ and well below the UV cutoff, one still expects higher-dimension operators to be suppressed with respect to the quartic one by powers of small couplings, since well below the UV cutoff the coefficients of higher-order operators should be dominated by "renormalizable" couplings. 

\begin{figure}[!h]
\includegraphics[width=\textwidth]{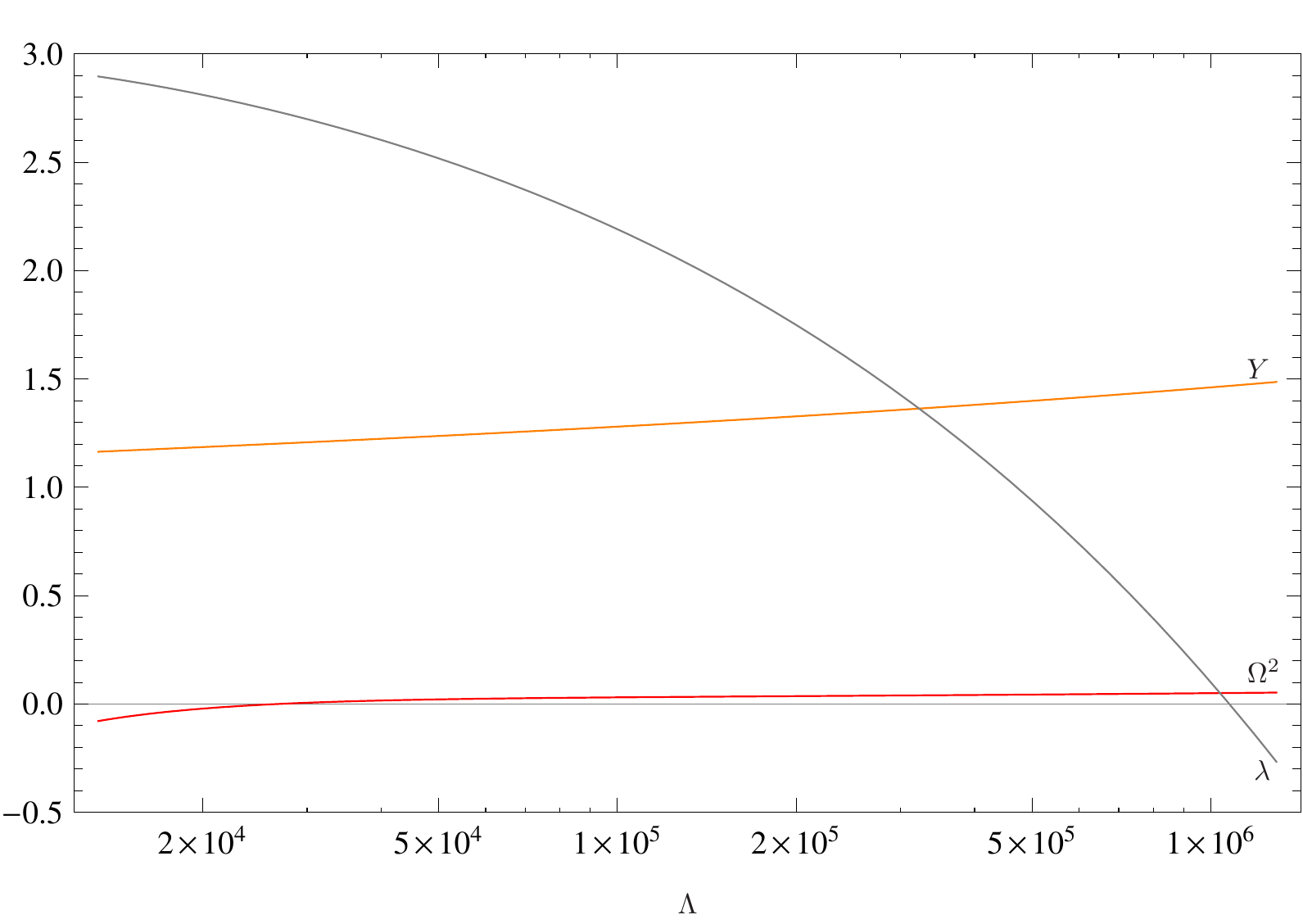}
\caption{Example of numerical solution of RGE in which radiative symmetry breaking takes place. Plot corresponds to values $Y_{\Lambda}=1.461$, $M_{\Lambda}^2= 5 \times 10^{10}$, $\lambda_{\Lambda}=0.1$ and $v_{\Lambda}={g}_{\Lambda}=m_{\Lambda}=0$ at $\Lambda=10^6$.\label{numerical_instability}}
\end{figure}

The example of a~solution demonstrating such features is plotted in the Fig. \ref{numerical_instability}. For this solution scalar mass parameter $\Omega_{\Lambda}^2$ vanishes at he scale $\Lambda=2.67 \times 10^4$ and quartic coupling $\lambda_{\Lambda}$ has a~zero at $\Lambda = 1.07 \times 10^6$. While investigating features of this solution one can notice a~strong dependence of the scale of symmetry breaking on the value of Yukawa coupling $Y_{\Lambda}$. This fine-tuning problem makes one choose very precisely the initial condition for Yukawa coupling in order to make the symmetry breaking scale low. 

\begin{figure}[!h]
\includegraphics[width=\textwidth]{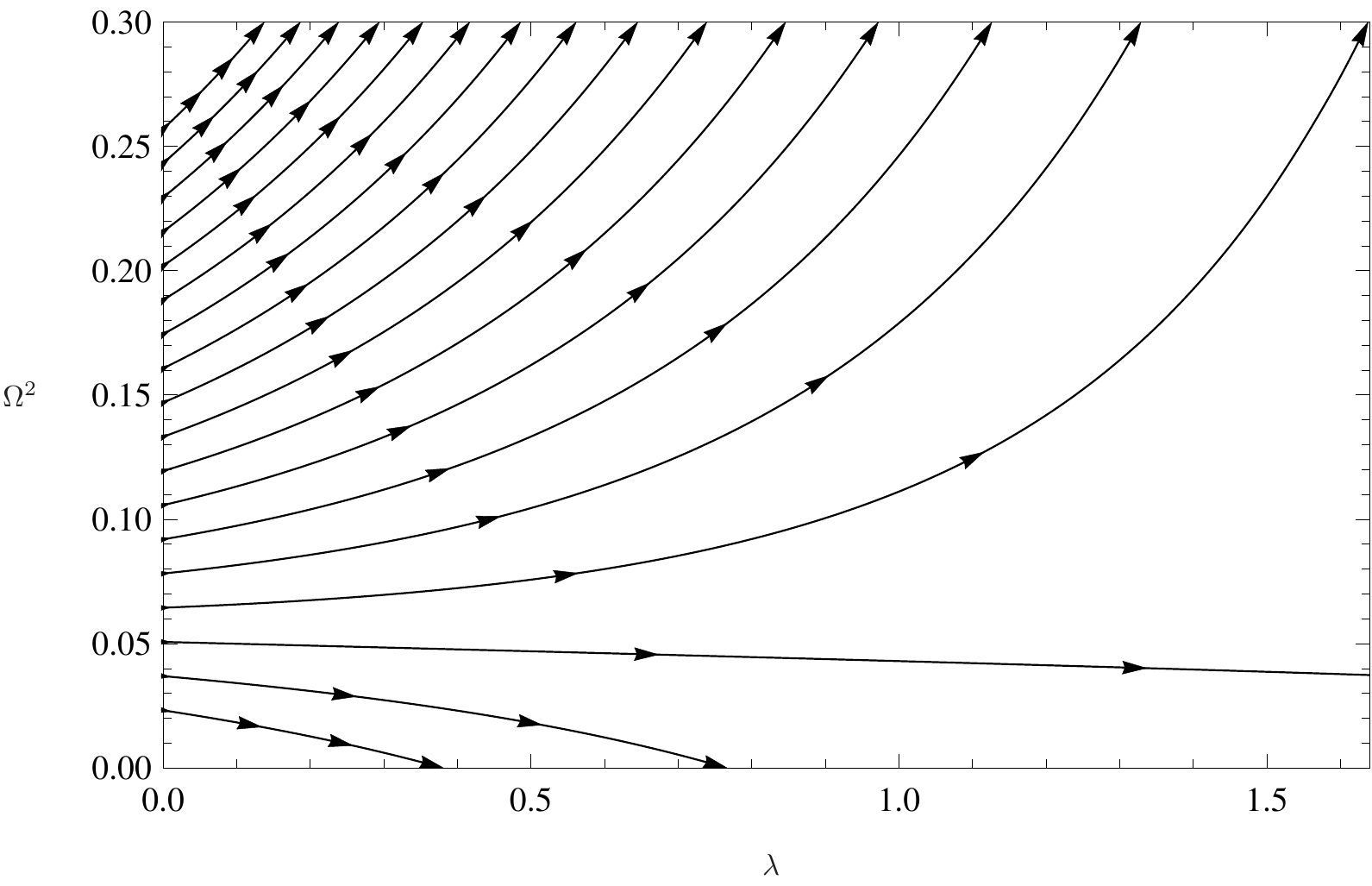}
\caption{Flow of $\lambda_{\Lambda}$ and $\Omega^2_{\Lambda}$ in the range of parameters where spontaneous symmetry breaking takes place. Plot corresponds to $Y_{\Lambda}=1.461$, $\lambda_{\Lambda}=0.1$, $v_{\Lambda}={g}_{\Lambda}=m_{\Lambda}=0$ and varying $M_{\Lambda}^2$ at $\Lambda=10^6$. Moving along the lines in the direction of arrows corresponds to decreasing scale \protect\nolinebreak[4]$\Lambda$.\label{numerical_Omegavslambda}}
\end{figure}

\begin{figure}[!h]
\includegraphics[width=\textwidth]{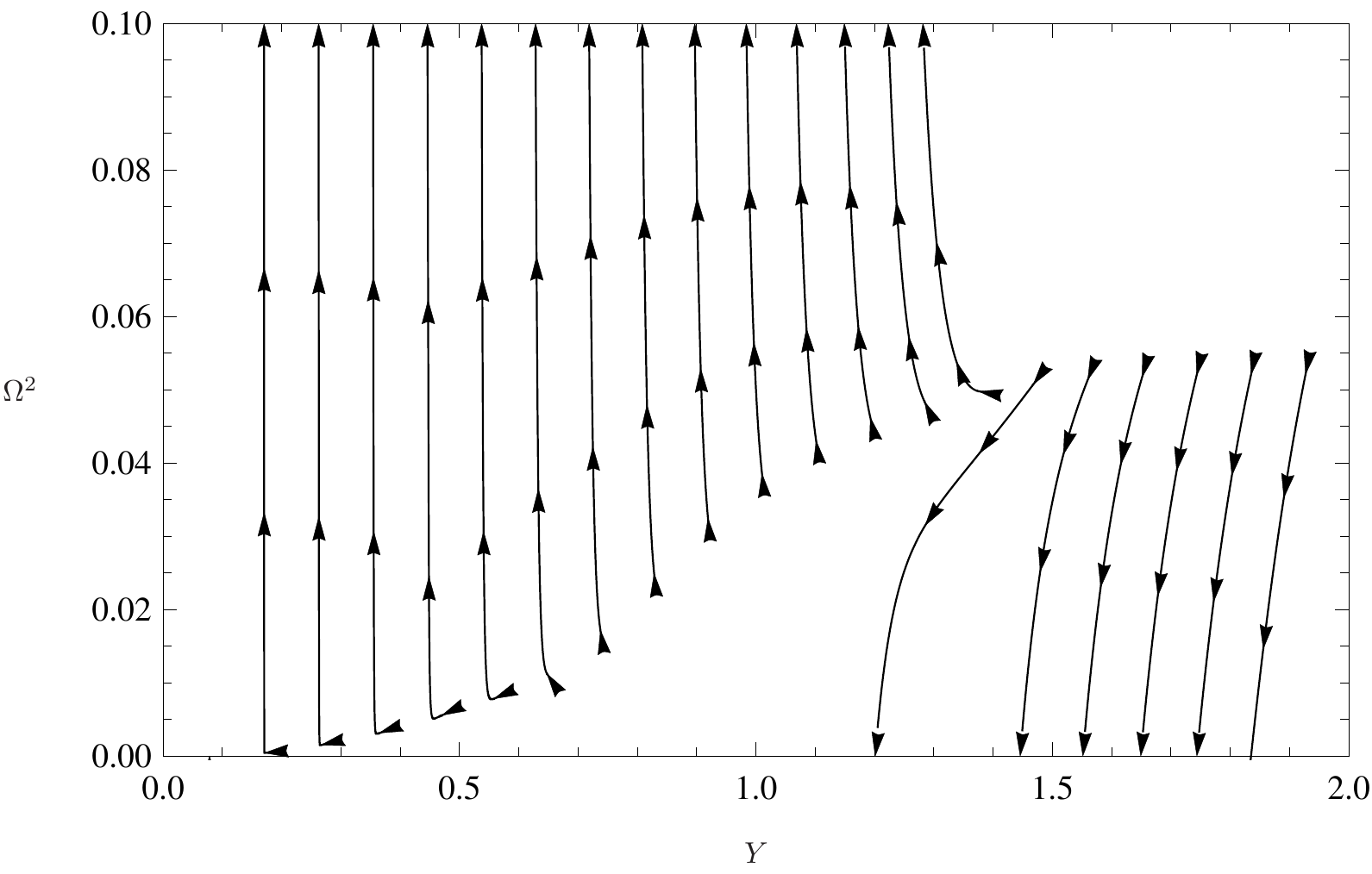}
\caption{Flow of $Y_{\Lambda}$ and $\Omega^2_{\Lambda}$ in the range of parameters where spontaneous symmetry breaking takes place. Plot corresponds to $\lambda_{\Lambda}=0.1$, $M_{\Lambda}^2= 5 \times 10^{10}$, $v_{\Lambda}={g}_{\Lambda}=m_{\Lambda}=0$ and varying $Y_{\Lambda}$ at $\Lambda=10^6$. Moving along the lines in the direction of arrows corresponds to decreasing scale \protect\nolinebreak[4]$\Lambda$.\label{numerical_OmegavsY}}
\end{figure}

The issue of spontaneous symmetry breaking can be studied with the help of the Fig. \ref{numerical_Omegavslambda} in which numerical solutions with different initial conditions are projected on the plane spanned by $\lambda_{\Lambda}$ and $\Omega^2_{\Lambda}$. All solutions given there have $Y_{\Lambda}=1.461$, $\lambda_{\Lambda}=0.1$ and $v_{\Lambda}={g}_{\Lambda}=m_{\Lambda}=0$ as a~initial conditions set at $\Lambda=10^6$, but initial value of $\Omega^2_{\Lambda}$ varying. From the Fig. \ref{numerical_Omegavslambda} one can see that if for any scale $\Lambda$ couplings will be lower than certain critical value $\Omega^2_{cr}$ then $\Omega^2$ will run negative in the IR and chiral parity will be spontaneously broken. Moreover as can be seen in the \mbox{Fig. \ref{numerical_OmegavsY}} critical value $\Omega^2_{cr}$ is rather sensitive to Yukawa coupling $Y$. From the behaviour shown in the Fig. \ref{numerical_OmegavsY} one concludes that $\Omega^2_{cr}$ decreases when the value of $Y$ increases and for any value of $\Omega^2$ there exist a~critical value of Yukawa coupling $Y_{cr}$. Once $Y_{cr}$ is exceeded, the radiative spontaneous symmetry breaking appears. Hence fourth quadrant of the Fig. \ref{numerical_OmegavsY} gives direct evidence of the Coleman--Weinberg mechanism at work. 

\section{Conclusions\label{conclusion}} 
In this paper we have used Wilsonian effective action to investigate fine-tuning and vacuum stability in a~simple model exhibiting spontaneous breaking of a~discrete symmetry and large fermionic radiative corrections which are able to destabilise quartic scalar self-coupling. 
Regulator independence of Wilsonian RG provides consistent and well-defined procedure to analyse the issue of quadratic divergences. In the simplified model simulating certain features of SM the Wilsonian renormalization group equations have been studied. We have explained in what sense the truncation adopted in the calculations corresponds to 1-loop \mbox{Gell-Mann--Low} running. In fact, in both cases the approximations used correspond to lowest-order quantum effects within each renormalization scheme. 

Numerical solutions of RGE have revealed interesting behaviour, caused by the same diagrams that generate quadratic divergences. An operator relevant near Gaussian fixed point (for example mass parameter for scalar particles) can run like marginal or even irrelevant operator, rather than decrease with growing scale. Furthermore solutions for different physical quantities flow close to each other with increasing scale. The flow in the direction of some common value indicates severe fine-tuning. In such a~situation small changes of boundary values of parameters at high scale produce very different vacuum expectation values for scalar field and other measurable quantities at low energies. We have estimated fine-tuning as a~function of scale of the effective theory. For all parameters the adopted measure of fine-tuning grows rapidly. Power-like functions fitted to the obtained fine-tuning curves grow faster than $\Lambda^2$ over the range of scales taken into account in the study. 

It should be stressed that Wilsonian RGEs, in contrast to \mbox{Gell-Mann--Low} running, accommodate automatically decoupling of heavy particles. As noticed in Section \ref{numerical}, the contribution to flow coming from particles with masses $M_{heavy}$ greater than the scale $\Lambda$ of the effective action is strongly suppressed. The main contributions to the interactions generated by heavy states are integrated out during calculation of the effective action for $\Lambda \ll M_{heavy}$ and are included in the effective Wilson coefficients. 
These properties of Wilsonian RGE explain why fine-tuning of Wilson parameters is so interesting. Let us imagine a~more fundamental theory (say theory A) in which the SM is embedded. If one calculates in theory A effective action for the scale $\Lambda$ below, but not very much, the lowest mass of the particles from the New Physics sector, one obtains certain values of the Wilson coefficients $c^{A}$. On the other hand one can extrapolate the flow obtained from the SM to the scale $\Lambda$ and calculate the Wilson coefficients $c^{SM}$. Couplings computed in both ways should match, that is 
$c^{A}_{\Lambda} =c^{SM}_{\Lambda}$. 
If the couplings $c^A$ are different from $c^{SM}$ at the level of fine-tuning $\Delta c$, that is $c^{A}_{\Lambda} (1 \pm \Delta c ) = c^{SM}_{\Lambda}$,
theory A will produce IR effective action completely different from the \nolinebreak[4]SM.

We have studied the issue of spontaneous symmetry breaking due to radiative corrections in the Wilsonian framework. We have demonstrated that there exists a~critical value $\Omega^2_{cr}$ below which $\Omega^2$ runs negative in the IR and symmetry becomes spontaneously broken. Moreover, critical value $\Omega^2_{cr}$ is sensitive to Yukawa coupling $Y$. One can see that $\Omega^2_{cr}$ decreases when the value of $Y$ increases and for any value of $\Omega^2$ there exist a~critical value of Yukawa coupling $Y_{cr}$. Once $Y_{cr}$ is exceeded, the radiative spontaneous symmetry breaking appears, which is a~direct evidence of Coleman--Weinberg mechanism at work. 
\pagebreak[4]
\begin{center}
{\bf Acknowledgements}
\end{center}
This work has been supported by National Science Center under research grant \linebreak[4]DEC-2012/04/A/ST2/00099\nolinebreak[4] and partially under research grant \linebreak[3]DEC-2011/01/M/ST2/02466\nolinebreak[4]. Authors (Z.L.) are grateful to the Mainz Institute for Theoretical Physics (MITP) for its hospitality and its partial support during the completion of this work.
\appendix
\section{Wilsonian RGE\label{Wilsonian}}
The aim of this section is to present brief introduction to the methods of Wilsonian RGE and to set the notation. We will define most of the terms that we used in the text of this paper. 
Method of discrete Renormalization Group Equations has been presented in Wilson's and Kogut's review \cite{Wilson:1973jj}, a~short introduction can be found in book of Peskin and Schroeder \cite{0201503972}. RGE for Legendre effective action have been derived in \cite{Wetterich:1992yh,Bonini:1992vh,Morris:1993qb}. Interesting formal developments on FRG (Functional Renormalisation Group) are presented in \cite{Rosten:2010vm}. Introductory review on FRG in gauge theories has been given in \cite{Gies:2006wv}. Further references can be found in the book \cite{Kopietz:2010zz} as well as in \cite{Polonyi:2001se} and \cite{Bagnuls:2000ae}.
\subsection{Wilsonian effective action\label{Wilsonian_effective_action}}
The fundamental object of QFT is generating functional $\mathcal{Z}[J]$ which is given in path integral formalism by formal expression\footnote{It is convenient to use Euclidean space in the context of Wilsonian RGE, so we will assume for a~moment that we use Euclidean formulation of QFT.}
\begin{equation}
\mathcal{Z}[J] = \int \mathcal{D} \Phi e^{-S_E[\Phi] + \int d^4 x_E J \Phi},\label{generating_functional}
\end{equation}
where $S_E$ is Euclidean action for the $\Phi$ field(s) and $J$ is source(s).
We can rewrite \eqref{generating_functional} in equivalent form \eqref{generating_functional_momentum} using Fourier transform\footnote{We will denote Fourier transform of field $F(x)$ as $\hat{F}(p)$.} $\hat{\Phi}$
\begin{equation}
\mathcal{Z}[\hat{J}] = \int \mathcal{D} \hat{\Phi} e^{-S_E[\hat{\Phi}] + \int \frac{d^4 p }{(2 \pi)^4} \hat{J} \hat{\Phi}} = \int \prod_p d \hat{\Phi} (p) e^{-S_E[\hat{\Phi}(p)] + \int \frac{d^4 p }{(2 \pi)^4} \hat{J} (p) \hat{\Phi}(p)}. \label{generating_functional_momentum}
\end{equation}
We introduce projection operator
\begin{equation}
\left( b_\Lambda \hat{F} \right) (p) = \theta_0 (p^2 - \Lambda^2) \hat{F} (p),
\end{equation} 
where $\theta_0$ is defined as follows
\begin{equation}
\theta_0 (x) = \left\{\begin{array}{ll} 1 & \textrm{for } x > 0 \\ 0 & \textrm{for } x \le 0\end{array} \right. . \label{Heaviside}
\end{equation}
We can divide integration variables $\hat{\Phi} (p)$ into two classes:
\begin{align}
\hat{\Phi}_{\le} (p) &= (1- b_\Lambda) \hat{\Phi}(p), & \hat{\Phi}_> (p)&=b_\Lambda \hat{\Phi}(p).
\end{align}
It is easy to find that
\begin{equation}
\prod_p d \hat{\Phi}(p)=\left(\prod_{p \colon p^2 \le \Lambda^2} d \hat{\Phi}_{\le}(p)\right) \left(\prod_{p' \colon p'^2 > \Lambda^2} d \hat{\Phi}_>(p')\right).
\end{equation}
Action $S_E[\hat{\Phi}]$ can be rewritten as
\begin{equation}
S_E[ \hat{\Phi}_{\le} + \hat{\Phi}_>] = S_E[ \hat{\Phi}_{\le}]+S_E[ \hat{\Phi}_>]+\mathcal{S}_E[ \hat{\Phi}_{\le},\hat{\Phi}_>],
\end{equation}
where we denoted by $\mathcal{S}_E[ \hat{\Phi}_{\le},\hat{\Phi}_>]$ the part which depends on both $ \hat{\Phi}_{\le}$ and $\hat{\Phi}_>$.
In this notation generating functional $\mathcal{Z}[\hat{J}]$ takes form
\begin{multline}
\mathcal{Z}[ \hat{J}_{\le},\hat{J}_>] = \int \!\!\! \prod_{p \colon p^2 \le \Lambda^2} d \hat{\Phi}_{\le}(p) e^{-S_E[ \hat{\Phi}_{\le}(p)] + \int \frac{d^4 p }{(2 \pi)^4} \hat{J}_{\le}(p) \hat{\Phi}_{\le}(p)}\\
\cdot \int \!\!\! \prod_{p' \colon p'^2 > \Lambda^2} d \hat{\Phi}_>(p') e^{-S_E[\hat{\Phi}_>(p') ]-\mathcal{S}_E[\hat{\Phi}_{\le}(p),\hat{\Phi}_>(p') ] + \int \frac{d^4 p }{(2 \pi)^4} \hat{J}_>(p') \hat{\Phi}_>(p') }
\end{multline}
where we have used orthogonality of modes with different momenta and have defined
\begin{align}
\hat{J}_{\le} (p) &= (1- b_\Lambda) \hat{J}(p), & \hat{J}_> (p)&=b_\Lambda \hat{J}(p).
\end{align}
Let us now imagine for a~moment that we integrate over all $\hat{\Phi}_>$ in generating functional $\mathcal{Z}[ \hat{J}_{\le},\hat{J}_>]$. Then we will obtain
\begin{equation}
\mathcal{Z}[ \hat{J}_{\le},\hat{J}_>] = \int \!\!\! \prod_{p \colon p^2 \le \Lambda^2} d \hat{\Phi}_{\le}(p) e^{-S_E[\hat{\Phi}_{\le}]+\int \frac{d^4 p }{(2 \pi)^4} \hat{J}_{\le}(p) \hat{\Phi}_{\le}(p)} \mathcal{Z}_> [\hat{J}_>].
\end{equation}
If we restricts ourselves to generating functional $\mathcal{Z}_{\le}[\hat{J}_{\le}]$ for Greens functions of low-energy modes $\hat{\Phi}_{\le}$ then it will get form:
\begin{equation}
\mathcal{Z}_{\le}[\hat{J}_{\le}] = \int \!\!\! \prod_{p \colon p^2 \le \Lambda^2} d \hat{\Phi}_{\le} (p) e^{-S_E[\hat{\Phi}_{\le}]+\log \mathcal{Z}_> [0] +\int \frac{d^4 p }{(2 \pi)^4} \hat{J}_{\le}(p) \hat{\Phi}_{\le}(p)}. \label{generating_functional_integrated}
\end{equation} 
Action $S_\Lambda [\hat{\Phi}_{\le}] := S_E[\hat{\Phi}_{\le}]-\log \mathcal{Z}_> [0]$ is called Wilsonian effective action for scale $\Lambda$.

\subsection{Flow equations\label{flow_equations}}
We define generating functional $W_\Lambda [\hat{J}]$ by equation
\begin{equation}
e^{W_\Lambda[\hat{J}]}:= \int \!\!\! \prod_{p \colon p^2 > \Lambda^2} d \hat{\Phi}_{>} (p) e^{-S_E[\hat{\Phi}_{\le}+\hat{\Phi}_>] +\int \frac{d^4 p }{(2 \pi)^4} \hat{J}(p) \left( \hat{\Phi}_{\le}(p) + \hat{\Phi}_{>}(p)\right)}.\label{W_definition}
\end{equation}
If $S_E$ can be decomposed as
\begin{equation}
S_E[\hat{\varphi}] =: S_{I}[\hat{\varphi}] + S_{G}[\hat{\varphi}] \label{interaction}
\end{equation}
with some arbitrary interaction part $S_{I}$, and a~Gaussian part of the general form
\begin{equation}
S_{G}[\hat{\varphi}]=:- \int \frac{d^4 p}{(2 \pi)^4} \frac{1}{2}\hat{\varphi}(-p) {G}^{-1} \hat{\varphi}(p). \label{Gaussian}
\end{equation}
then it can be showed that following definition
\begin{equation}
e^{W_\Lambda[\hat{J}]} = \int \prod_p d \hat{\Phi} (p) e^{-S_I[\hat{\Phi}(p)] + \int \frac{d^4 p}{(2 \pi)^4}\left[ \frac{1}{2} \hat{\Phi} (-p) \theta_0 (p^2-\Lambda^2) {G}^{-1}(p) \hat{\Phi}(p) + \hat{J} (p) \hat{\Phi}(p)\right]}. \label{effective_partition}
\end{equation}
is equivalent to \eqref{W_definition}.

For calculating $\beta$-functions, it is convenient to consider Legendre transform $\Gamma_\Lambda[\phi]$ of generating functional $W_\Lambda[J]$.
Before we define Wilsonian RGE we should reduce redundant degrees of freedom in $\Gamma_\Lambda [\phi]$. Firstly it can happen that
\begin{equation}
\left.\frac{\delta \Gamma_\Lambda}{\delta \phi} \right|_{\phi=0} \neq 0.
\end{equation}
In such a~case we shift the the field $\phi\mapsto \phi_0 + \varphi$ by the solution $\phi_0$ of equation $\left.\frac{\delta \Gamma_\Lambda}{\delta \phi} \right|_{\phi=\phi_0} = 0$. Secondly there are redundant degree of freedom corresponding to the rescaling of the field $\varphi$. We assume that $\varphi$ has canonical kinetic term in $\Gamma_\Lambda[\varphi]$. Wilsonian RGE are differential equations describing change of $\Gamma_\Lambda[\varphi]$ due to the change of scale $\Lambda$.

Legendre effective action $\Gamma_\Lambda[\varphi]$ can be expanded in Taylor series in powers of field $\varphi$
\begin{equation}
\Gamma_\Lambda[\hat{\varphi}] = \sum_{n \in \mathbb{N}} \frac{1}{n!} \hat{\varphi}^n \frac{\delta^n}{\delta \hat{\varphi}^n} \Gamma_\Lambda.
\end{equation}
Each coefficient of Taylor expansion can be expanded in derivatives of $\varphi$ 
\begin{equation}
\frac{\delta}{\delta \hat{\varphi}(p_1)} \cdots \frac{\delta}{\delta \hat{\varphi}(p_n)} \Gamma_\Lambda= \left(\prod_{i=1}^n \int \frac{d^4 p_i}{(2 \pi)^4}\right) \delta \left(\sum_{i=1}^n p_i\right) \sum_{i_1, \dots i_n \in \mathbb{N}} C^{(n)}_{i_1, \dots, i_n} p_1^{i_1} \cdots p_n^{i_n}. 
\end{equation}
Coefficients $C^{(n)}_i$ are called Wilson coefficients. It is convenient to use dimensionless parameters
\begin{equation}
c^{(n)}_i =C^{(n)}_i \Lambda^{-\dim C^{(n)}_i}
\end{equation}
where $\dim C^{(n)}_i$ is canonical dimension of the coefficient. RGE expressed in terms of dimensionless parameters is dynamical system.

Legendre effective action $\Gamma_\Lambda[\varphi]$ in the limit $\Lambda \to \infty$ is called classical action. This limit usually does not exist in strict mathematical sense, because Wilson coefficients are typically divergent when $\Lambda \to \infty$. Classical action for theories that are not finite should be thought as a~formal expression which must be renormalized.

\subsection{Perturbation theory}
Wilsonian effective action and flow equations can be approximately derived in perturbation theory and expressed by Feynman diagrams. Starting with \eqref{effective_partition} the Wilsonian effective action $W_\Lambda[\hat{J}]$ (so Legendre effective action $\Gamma_\Lambda[\varphi]$ too) can be computed in perturbation theory in a~manner analogous to computation of the generating functional for connected Green's functions $W[\hat{J}]$ (1PI effective action $\Gamma_{\textrm{1PI}}[\varphi]$), but with a~propagator substituted by a~propagator with a~cutoff:
\begin{equation}
G_\Lambda (p) := \theta_0 (p^2-\Lambda^2) G (p).
\end{equation}
Legendre effective action $\Gamma_\Lambda[\varphi]$, like the 1PI effective action $\Gamma_{\textrm{1PI}}[\varphi]$, has usually infinitely many terms. Wilsonian RGE for a~generic theory is the set of infinitely many coupled equations for infinitely many couplings which rarely can be solved exactly. For practical purposes one typically have to use certain approximation strategy. The most common consists in truncation of effective action and in considering only a~subset of Wilson coefficients which are relevant for the problem in question.\section{Derivatives of loop integrals with IR cutoff\label{loop_integrals}}
During calculation of RGE we used following derivatives of integrals $I_N(R)$ defined in \cite{Bilal:2007ne}.
\begin{equation}
\frac{\partial}{\partial \Lambda} I_1(R) = \Lambda^2 \frac{1}{\Lambda} \frac{-2i}{(4 \pi)^2} \frac{1}{1+\frac{R}{\Lambda^2}}+\BigO{\varepsilon}
\end{equation}
\begin{equation}
\frac{\partial}{\partial \Lambda} I_2(R) = \frac{1}{\Lambda} \frac{-2i}{(4 \pi)^2} \frac{1}{(1+\frac{R}{\Lambda^2})^2}+\BigO{\varepsilon}
\end{equation}
\begin{equation}
\frac{\partial}{\partial \Lambda} I_N(R) = \Lambda^{-2(N-2)} \frac{1}{\Lambda} \frac{-2i}{(4 \pi)^2}\frac{1}{(1+\frac{R}{\Lambda^2})^N}+\BigO{\varepsilon}
\end{equation}
\section{1-loop matching conditions\label{matching}}
We used 1-loop matching conditions in order to express Wilsonian coefficients of the effective action at the matching scale $\Lambda$ by physical quantities. The later can be formally obtained from the effective action in the limit of the scale  $\Lambda \to 0$, which however is out of reach of numerical methods as discussed in Section \ref{fine_tuning}.

FeynArts/FeynCalc packages for Mathematica were employed to calculate matching conditions. In addition we used ANT package modified to become compatible with FeynCalc. Moreover, we added infinite terms listed in \cite{Angel:2013hla} to the expressions in this package in order to check consistency of our calculation. We have written an analogous package that contains loop integrals in dimensional regularization with IR cutoff.

We expressed parameters of the effective action in terms of physical vacuum expectation value $v_{ph}$ of $\Phi$ field, pole masses $m_{ph}$, $M_{ph}$ of respectively $\Psi$ and $\Phi$ fields. We have defined Yukawa coupling $Y_{ph}$ and trilinear coupling ${g}_{ph}$ as a~1PI part of scattering amplitude at kinematic point with incoming momentum square equal to $\mu^2$ and outgoing momenta squares equal to $\mu^2$ and $0$. In definition of $Y_{ph}$ we assumed that outgoing fermion has non-zero momentum. Coupling $\lambda_{ph}$ is defined at kinematic point with Mandelstam variables equal to $\mu^2$.

Diagrams that contribute to matching conditions are discussed in Section \ref{RGE} and are presented in the Figs. from \ref{v_broken} to \ref{lambda_broken}.

\bibliographystyle{elsarticle-num}
\bibliography{FTVSWEA}
\end{document}